\documentclass[aps,prb,onecolumn,superscriptaddress,amsmath,amssymb]{revtex4-2}

\usepackage[utf8]{inputenc}
\usepackage{amsmath,amsthm}
\usepackage{graphicx}
\usepackage{tikz}
\usepackage{tikz-3dplot}
\usetikzlibrary{shadows}
\usetikzlibrary{positioning}
\usetikzlibrary{arrows.meta}
\usepackage[colorlinks = true,
            linkcolor = blue,
            urlcolor  = blue  ,
            citecolor = blue,
            anchorcolor = blue]{hyperref}
\usepackage{braket}
\usepackage{xcolor}

\newcommand{\x}{{x}}

\newcommand{\bea}{\begin{eqnarray}}
\newcommand{\eea}{\end{eqnarray}}

 \global\long\def\de{\delta}
 
\global\long\def\th{\theta}
\global\long\def\th{\theta}

\global\long\def\ell#1{\theta_{#1}}

\global\long\def\si{\sigma}

\global\long\def\s{\sigma}

\global\long\def\eps{\epsilon}
\global\long\def\al{\alpha}

 \global\long\def\de{\delta}

\global\long\def\no{\nonumber}

\newtheorem{cor}{Corollary}

\theoremstyle{remark}
\newtheorem*{rem}{Remark}

\newcommand{\E}{{\cal E}}
\newcommand{\pu}{{p}}
\newcommand{\uu}{{u}}
\global\long\def\braket#1#2{\left\langle #1|#2\right\rangle }

\newcommand{\eE}{\mathrm{e}}

\def\i{{\rm i}}

\theoremstyle{plain}

\def\epc{,}
\def\epp{.}
\let\p=\pi

\usepackage{amsfonts}
\usepackage{amsmath}
\usepackage{amsthm}
\usepackage{amssymb}
\usepackage{amscd}
\usepackage{eucal}
\usepackage{bm}
\usepackage{graphicx}

\newcommand{\lvp}{\bm{k}}
\newcommand{\lp}{k}
\newcommand{\vvu}{\bm{u}}
\newcommand{\vu}{u}
\newcommand{\vupb}{\bm{p}}
\newcommand{\vup}{p}
\newcommand{\JJ}{}
\newcommand{\currJ}{J}
\newcommand{\ik}{l}
\newcommand{\nvn}{\bm{n}}
\newcommand{\barT}{T}

\def\nn{\mathbf{n}}

\newcommand{\A}{\mathcal{A}}

\newcommand{\ir}{\mathrm{i}}

\input{macro_xin.lib}

\allowdisplaybreaks

\begin{document}

\title{Chiral eigenbases of the XX and XY quantum spin chains}
\bigskip

\author{Xin  Zhang}
\affiliation{Beijing National Laboratory for Condensed Matter Physics, Institute of Physics, Chinese Academy of Sciences, Beijing 100190, China}
\author{  Frank G\"ohmann} 
\affiliation{Department of Physics,
 University of Wuppertal, Gaussstra\ss e 20, 42119 Wuppertal,
  Germany}
\author{  Andreas Kl\"umper} 
\affiliation{Department of Physics,
 University of Wuppertal, Gaussstra\ss e 20, 42119 Wuppertal,
  Germany}
\author{Vladislav Popkov}
 \affiliation{Department of Physics,
  University of Wuppertal, Gaussstra\ss e 20, 42119 Wuppertal,
  Germany}
\affiliation{Faculty of Mathematics and Physics, University of Ljubljana, Jadranska 19, SI-1000 Ljubljana, Slovenia}
 
\begin{abstract}
We calculate the values of observables in chiral eigenstates
of the XX quantum spin chain that were introduced in 
previous work and compare the form of the result
with the respective expressions obtained in the more familiar
eigenbasis of states with fixed magnetization in $z$-direction.
We carry out the diagonalization of the XY spin chain in the
chiral basis. We calculate the norm of the chiral XY eigenstates,
and the values of the one-point functions and some neighbor
two-point correlation functions. We interpret the spectrum and
the particle content of the XY chain in terms of scattering
states of an even number of kink and anti-kink excitations
that are created over a reduced Brillouin zone.
\end{abstract}
\maketitle

\textbf{Introduction--} 
Chiral quantum states with nontrivial topology are becoming an important
tool in quantum technologies. With atoms in optical lattices it is possible 
to pass from non-interacting to strongly correlated spin states by 
adiabatic manipulations in external electromagnetic fields 
\cite{2023KetterleRotationArXiv, 2020KetterleDaley,2020VenegasGomez}.
The nontrivial topology of chiral eigenstates entails their exceptional
robustness against noise even compared to the ground state \cite{ChiralMoreStable}.  
It appears therefore natural to describe generic chiral states (not
necessarily eigenstates) and their time evolution not in terms of the
conventional computational basis (which is topologically trivial), but
rather in terms of a basis which is chiral and topologically nontrivial
by itself. A chiral basis for qubits was proposed in \cite{2024ChiralBasis},
where it was employed to study the relaxation of spin helix states in the XX
chain. This chiral basis shares all the convenient features of the computational
basis, such as the product structure and orthonormality, but at the same
time is intrinsically topological, as will be seen below. 
 
The convenience of the chiral basis can be illustrated with a simple
example: let us take a spin helix state 
\begin{align} \label{def:SHS}
     \ket{\Phi(\al,\eta)} = \bigotimes_n \frac{1}{\sqrt{2}}\binom{1}{\eE^{\i(\alpha+n\eta)}},\quad \alpha,\eta\in\mathbb{R},
\end{align}
where $\binom{1}{\eE^{\ir v}}$ describes the state of a qubit, while $+n\eta$
represents a linear increase of the qubit phase along the chain.
The qubit polarization along the chain forms a helix of period $2\pi /\eta$
in the $xy$-plane. A shift of the initial phase $\al$ in (\ref{def:SHS})
does not affect the physical properties of the helix.
However, representing such a phase shift in the computational basis
is inconvenient: it will affect all the terms of the expansion in
different manners. On the contrary, the chiral qubit basis \cite{2024ChiralBasis}
allows to accomodate the phase $\al$ by a global rotation about the
$z$-axis.

In this work we intend to demonstrate that the practical usage of a chiral
qubit basis can be as simple as the usage of the computational basis of
$S^z$ eigenstates. This is true, in particular, in connection with the XX
quantum spin chain \cite{LSM61} that can be diagonalized in either basis as
the Hamiltonian commutes with the operator of the $z$-component of the total spin $S^z$
as well as with a winding number operator $V$ \cite{2024ChiralBasis}
(see (\ref{eq:Voper}) below) that measures the chirality in the $xy$-plane.
This means that the XX Hamiltonian is block diagonal in both bases, with
blocks that either have fixed values of the total magnetization in
$z$-direction or fixed winding numbers in the $xy$-plane. Both block
structures are, however, incompatible as $S^z$ and $V$ do not commute.
For this reason, the magnetization in $z$-direction and the winding
number cannot be measured simultaneously.

In both eigenbases the eigenfunctions of the XX Hamiltonian are
parametrized in terms of quasimomenta, or Bethe roots, that satisfy
simple Bethe equations. Although the Bethe equations are of similar
form in both cases, the meaning of the usual quasimomenta $\lp_j$ in
the $S^z$ eigenbasis and the quasimomenta $p_j$ in the chiral basis
is different. In particular, simple spectral observables and expectation
values of one-point and neighbor two-point correlation functions are not
necessarily expressed the same way in both types of quasimomenta.
We shall discuss a number of examples below.

In the second part of our manuscript we diagonalize the anisotropic XY
Hamiltonian $H_{XY}$ in the chiral basis. $H_{XY}$ is not rotation invariant
and does not commute with $S^z$. Still, it commutes with the winding number
operator $V$ that provides a $U(1)$ symmetry and a block structure in this
case. The XY model was introduced and solved, by mapping it to a model of
non-interacting Fermions, in the seminal work \cite{LSM61} of Lieb, Schultz
and Mattis. For this model Bethe Ansatz solutions other than those that
can be obtained as special cases from Bethe Ansatz solutions of the more
general XYZ chain are not available.  For the XYZ chain a coordinate Bethe
Ansatz solution was obtained by Baxter \cite{Baxter73i,Baxter73ii,Baxter73iii},
and algebraic versions are due to Tahktadjan and Faddeev \cite{TaFa79}
(see also \cite{2020SlavnovXY,2023SlavnovXY}) and to Felder and Varchenko
\cite{FeVa96}. These solutions have in common that they diagonalize the
transfer matrix of the underlying eight-vertex model which does not commute
with the winding number operator $V$. Recently, three of the authors proposed
a new variant of the coordinate Bethe Ansatz for the XYZ chain
\cite{2024PedestrianXYZ} that applies to the XYZ Hamiltonian, but not to the
transfer matrix.  As we shall see, the XY limit of this Bethe Ansatz solution
gives a chiral orthonormal eigenbasis of the model. The eigenfunctions inherit
their parametrization in terms of rapidity variables from the XYZ chain. They
take the form of determinants with entries that are elliptic functions. We
also provide a reparametrization in terms of quasimomenta. In terms of these
the eigenfunctions take a particularly simple form and allow us to interpret
the elementary excitation of the finite chain as kink-anti-kink pairs
living in a reduced Brillouin zone. We finally provide explicit expressions of
some simple observables in terms of the quasimomenta.

\section{\boldmath XX model: $S^z$ versus chiral eigenbasis}
The one-dimensional periodic spin-$\frac12$ XX model is defined by the Hamiltonian
\begin{align} \label{def:XX0}
     H_{XX} = \JJ  \sum_{n=1}^{N} \left(\si_n^x \si_{n+1}^x + \si_n^y \si_{n+1}^y \right),
                \quad \si_{N+1}^\al \equiv \si_1^\al \epc
\end{align}
where $\si_n^\al$, $\al = x, y, z$, act as Pauli matrices on
the $n$th spin in the chain and where we set the exchange parameter equal to 1.
The XX model is among the simplest exactly solvable many-body systems.
Its solvability can be attributed to the fact that it can be mapped
to a model of non-interacting Fermions by a Jordan-Wigner transformation
\cite{LSM61}. Alternatively, the model can be related to the six-vertex
model \cite{Sutherland67,Baxter71a} at the free Fermion point. Within
this interpretation the Hamiltonian is proportional to the logarithmic
derivative of the six-vertex model transfer matrix. Its spectrum and
eigenfunctions can therefore be obtained by means of the algebraic
Bethe Ansatz \cite{STF79}. For the application of the latter method
it is crucial that the transfer matrix (and hence the Hamiltonian)
commutes with the operator
\begin{equation}
     S^z = \frac{1}{2} \sum_{n=1}^N \si_n^z
\end{equation}
of the $z$-component of the total spin.

Let $|\auf\,\>_z = {1 \choose 0}$ and $|\ab\,\>_z = {0 \choose 1}$
be normalized eigenvectors of $\s^z$ with eigenvalues $\pm 1$.
Then the states
\begin{equation} \label{szbasis}
     |s_1 s_2 \dots s_N\>_z =
        |s_1\>_z \otimes |s_2\>_z \otimes \dots \otimes |s_N\>_z
\end{equation}
with $s_j \in \{\auf, \ab\}$ form a basis of eigenstates of $S^z$ on
the space of states ${\cal H}_N = {({\mathbb C}^2)}^{\otimes N}$ of the
XX Hamiltonian (\ref{def:XX0}). Since $[H_{XX},\, S^z] = 0$, the Hamiltonian
(\ref{def:XX0}) is block diagonal in this basis, and the blocks
can be labeled by the eigenvalues $-N/2, -N/2 + 1, \dots, N/2$ of
$S^z$. A state with $m$ $\auf$-spins and $N - m$
$\ab$-spins corresponds to an $S^z$ eigenvalue of $m - N/2$.
Clearly, the number of such states is $N \choose m$. An efficient
way to address these states is to specify the positions of $\auf$-spins
on a background of $\ab$-spins. For this purpose choose 
$m$ integers $n_j$, $1 \le n_1 < n_2 < \dots < n_m\leq N$ and set
$\nvn = (n_1, \dots, n_m)$ and $|\nvn\>_z = \s_{n_1}^+ \dots \s_{n_m}^+
|\ab \dots \ab\>_z$. The entirety of such states for fixed $m$ is
a basis of the $\langle S^z\rangle = m - N/2$ subspace of ${\cal H}_N$.

\begin{theorem}\label{thm1}
[See e.g.\ \cite{1993ColomoIzergin}].
For $m=1,2,\ldots,N$, define $\lvp=(\lp_1,\lp_2,\ldots,\lp_m)$, $\lp_j
\in (-\pi,\pi]$ with $\lp_1 < \dots < \lp_m$.
The states
\begin{equation}
     |\xi_{m}(\lvp)\rangle
        = \sum_{1 \le n_1 < \dots < n_m \le N}\chi_{\nvn} (\lvp) |\nvn\>_z,\quad
     \chi_{\nvn} (\lvp) = \frac{1}{\sqrt{N^M}} \,
        \det_{j,\ik = 1,\dots,m}\bigl\{\re^{\i \lp_j n_\ik}\bigr\},
\label{XXEigenstatesConventional}
\end{equation}
where the quasi-momenta $\lp_j$ satisfy the Bethe equations
$\eE^{\i N\lp_j}=(-1)^{m-1}$, together with the pseudo vacuum
$|\ab \dots \ab\>_z$, form a complete set of mutually orthogonal,
normalized eigenstates of the Hamiltonian (\ref{def:XX0}). The
corresponding energy eigenvalues are
\begin{equation}
     E(\lvp)= 4\JJ  \sum_{j=1}^{m} \cos\lp_j.
\end{equation}
\end{theorem}

We call the basis $\{|\xi_{m}(\lvp)\rangle\}$ the $S^z$ eigenbasis
of the XX Hamiltonian (\ref{def:XX0}). It is of appealing simplicity
as the wave function within the blocks of fixed $S^z$ eigenvalues
take the form of Slater determinants.

In Ref. \cite{2024ChiralBasis} we constructed another rather different eigenbasis
of the XX Hamiltonian, the chiral eigenbasis, which is almost as simple
as the $S^z$ eigenbasis. The chiral basis is a basis of joint eigenvectors
of $H_{XX}$ and the operator
\begin{align}
&V = \frac{1}{4} \sum_{\ik=1}^{N/2} \left(  \si_{2\ik-1}^x   \si_{2\ik}^y
- \si_{2\ik}^y   \si_{2\ik+1}^x
\right),
\label{eq:Voper}
\end{align}
whose commutativity with $H_{XX}$ follows from its commutativity with
the more general XY Hamiltonian which is proved in App.~\ref{app:winding}.
Like the eigenstates of $S^z$, all eigenstates of $V$ can be chosen
as tensor products of eigenstates of Pauli matrices. All states
\begin{equation} \label{cpdstate}
|\kappa; \nvn\rangle = (-\i)^{\sum_{j=1}^M n_j} \bigotimes_{\ik=1}^{n_1}
                  \psi(\ik-\kappa)\bigotimes_{\ik=n_1+1}^{n_2} \psi(\ik-2-\kappa) \cdots
		  \bigotimes_{\ik = n_M + 1}^{N} \psi(\ik-2M-\kappa), \qquad
                  \psi(u) = \frac{1}{\sqrt{2}}\binom {1}{\i^{u}},
\end{equation}
where $\kappa=\pm1$, $\nvn = (n_1, \dots, n_M)$, $1 \le n_1 < n_2 <
\dots < n_M \le N$, are eigenstates of $V$ and form an orthonormal
basis of ${\cal H}_N$ which we call the chiral basis. In such states
the qubit polarization changes from factor to factor by an angle of
$+\pi/2$ or $-\pi/2$ in the $xy$-plane. Each anticlockwise or clockwise
rotation by $\pi/2$ adds $+\frac{1}{4}$ or $-\frac{1}{4}$ to the
eigenvalue of $V$ so that  
\begin{equation}
  V |\kappa;\nvn\rangle
  =\tst{\4} (N-2M) |\kappa;\nvn\rangle,\quad \kappa=\pm1
 \epc
\end{equation}
where $M$ is the number of clockwise rotations, further referred to
as \textit{kinks}. We say a kink is located at $n_j$ if the rotation
between site $n_j$ and site $n_{j+1}$ is clockwise. Each state
(\ref{cpdstate}) is uniquely characterized by its kink positions $\nvn$ and
the polarization of the first qubit, where $\k = +1, -1$ correspond to the
eigenstates of $\sigma^x$ with eigenvalues $+1$ and $-1$ respectively.

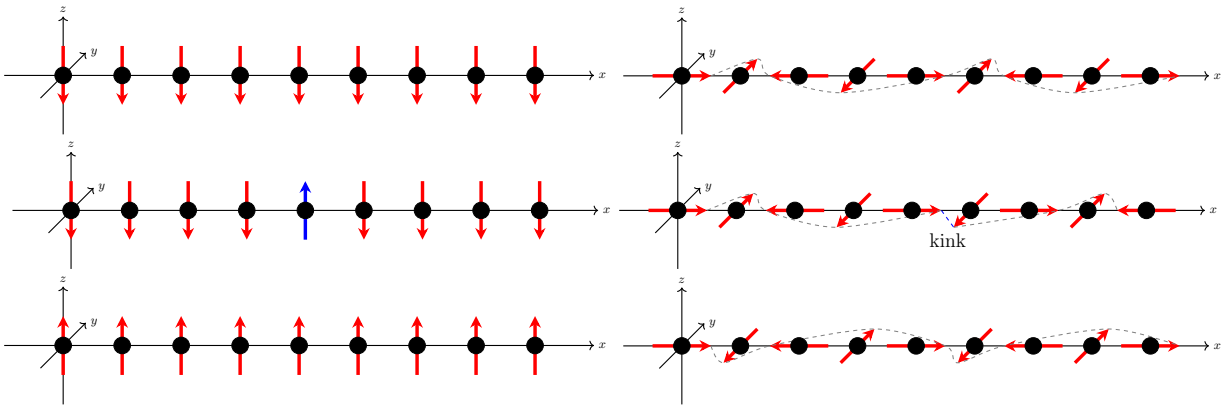
\begin{figure}
\centering
\resizebox{0.45\textwidth}{!}{\begin{tikzpicture}[scale=1.5]
\draw[thick,->] (-1,0,0) -- (9,0,0) node[anchor=west] {$x$};
\draw[thick,->] (0,-1,0) -- (0,1,0) node[anchor=south] {$z$};
\draw[thick,->] (0,0,1) -- (0,0,-1) node[anchor=west] {$y$};
\foreach \x in {0,...,8}
\draw[color=red,-{Stealth[length=3mm,width=3mm]},line width=2.5pt]
 (\x,0.5,0)--(\x,-0.5,0);
\foreach \x in {0,...,8}
\draw[fill=black] (\x,0,0) circle (4pt);
\end{tikzpicture}}
\resizebox{0.45\textwidth}{!}{
\begin{tikzpicture}[scale=1.5]
\draw[thick,->] (-1,0,0) -- (9,0,0) node[anchor=west] {$x$};
\draw[thick,->] (0,-1,0) -- (0,1,0) node[anchor=south] {$z$};
\draw[thick,->] (0,0,1) -- (0,0,-1) node[anchor=west] {$y$};
\draw[color=red,-{Stealth[length=3mm,width=3mm]},line width=2.5pt]
 (-0.5,0,0)--(0.5,0,0);
\draw[color=red,-{Stealth[length=3mm,width=3mm]},line width=2.5pt]
 (1,0,0.75)--(1,0,-0.75);
\draw[color=red,-{Stealth[length=3mm,width=3mm]},line width=2.5pt]
 (2.5,0,0)--(2-0.5,0,0);
\draw[color=red,-{Stealth[length=3mm,width=3mm]},line width=2.5pt]
 (3,0,-0.75)--(3,0,0.75);
\draw[color=red,-{Stealth[length=3mm,width=3mm]},line width=2.5pt]
(4-0.5,0,0)--(4+0.5,0,0);
\draw[color=red,-{Stealth[length=3mm,width=3mm]},line width=2.5pt]
 (4+1,0,0.75)--(4+1,0,-0.75);
\draw[color=red,-{Stealth[length=3mm,width=3mm]},line width=2.5pt]
 (4+2.5,0,0)--(4+2-0.5,0,0);
\draw[color=red,-{Stealth[length=3mm,width=3mm]},,line width=2.5pt]
 (4+3,0,-0.75)--(4+3,0,0.75);
\draw[color=red,-{Stealth[length=3mm,width=3mm]},line width=2.5pt]
 (7.5,0,0)--(8.5,0,0);
\draw[smooth,dashed,gray,line width=0.75pt]
 plot coordinates {
  (0.5,0,0)
  (1,0,-0.75)
  (2-0.5,0,0)
  (3,0,0.75)
  (4.5,0,0)
  (5,0,-0.75)
  (5.5,0,0)
  (7,0,0.75)
  (8.5,0,0)
        };
\foreach \x in {0,...,8}
\draw[fill=black] (\x,0,0) circle (4pt);
\end{tikzpicture}}
\resizebox{0.45\textwidth}{!}{
\begin{tikzpicture}[scale=1.5]
\draw[thick,->] (-1,0,0) -- (9,0,0) node[anchor=west] {$x$};
\draw[thick,->] (0,-1,0) -- (0,1,0) node[anchor=south] {$z$};
\draw[thick,->] (0,0,1) -- (0,0,-1) node[anchor=west] {$y$};
\foreach \x in {0,...,3,5,6,7,8}
\draw[color=red,-{Stealth[length=3mm,width=3mm]},line width=2.5pt]
 (\x,0.5,0)--(\x,-0.5,0);
\draw[color=blue,-{Stealth[length=3mm,width=3mm]},line width=2.5pt]
 (4,-0.5,0)--(4,0.5,0);
\foreach \x in {0,...,8}
\draw[fill=black] (\x,0,0) circle (4pt);
\end{tikzpicture}}\resizebox{0.45\textwidth}{!}{
\begin{tikzpicture}[scale=1.5]
\draw[thick,->] (-1,0,0) -- (9,0,0) node[anchor=west] {$x$};
\draw[thick,->] (0,-1,0) -- (0,1,0) node[anchor=south] {$z$};
\draw[thick,->] (0,0,1) -- (0,0,-1) node[anchor=west] {$y$};
\draw[color=red,-{Stealth[length=3mm,width=3mm]},line width=2.5pt]
 (-0.5,0,0)--(0.5,0,0);
\draw[color=red,-{Stealth[length=3mm,width=3mm]},line width=2.5pt]
 (1,0,0.75)--(1,0,-0.75);
\draw[color=red,-{Stealth[length=3mm,width=3mm]},line width=2.5pt]
 (2.5,0,0)--(1.5,0,0);
\draw[color=red,-{Stealth[length=3mm,width=3mm]},line width=2.5pt]
 (3,0,-0.75)--(3,0,0.75);
\draw[color=red,-{Stealth[length=3mm,width=3mm]},line width=2.5pt]
(3.5,0,0)--(4.5,0,0);
\draw[color=red,-{Stealth[length=3mm,width=3mm]},line width=2.5pt]
 (5,0,-0.75)--(5,0,+0.75);
\draw[color=red,-{Stealth[length=3mm,width=3mm]},line width=2.5pt]
 (5.5,0,0)--(6.5,0,0);
\draw[color=red,-{Stealth[length=3mm,width=3mm]},,line width=2.5pt]
 (7,0,0.75)--(7,0,-0.75);
\draw[color=red,-{Stealth[length=3mm,width=3mm]},line width=2.5pt]
 (8.5,0,0)--(7.5,0,0);
\draw[smooth,dashed,gray,line width=0.75pt]
 plot coordinates {
  (0.5,0,0)
  (1,0,-0.75)
  (1.5,0,0)
  (3,0,0.75)
  (4.5,0,0)
        };
\draw[smooth,dashed,gray,line width=0.75pt]
 plot coordinates {
(5,0,0.75)
(6.5,0,0)
(7,0,-0.75)
(7.5,0,0)
        };
\draw[smooth,dashed,blue,line width=0.75pt]
 plot coordinates {
(4.5,0,0)
(5,0,0.75)
        };\coordinate[label=left:{\Large kink}] (a) at (5,-0.5,0);
\foreach \x in {0,...,8}
\draw[fill=black] (\x,0,0) circle (4pt);
\end{tikzpicture}}
\resizebox{0.45\textwidth}{!}{\begin{tikzpicture}[scale=1.5]
\draw[thick,->] (-1,0,0) -- (9,0,0) node[anchor=west] {$x$};
\draw[thick,->] (0,-1,0) -- (0,1,0) node[anchor=south] {$z$};
\draw[thick,->] (0,0,1) -- (0,0,-1) node[anchor=west] {$y$};
\foreach \x in {0,...,8}
\draw[color=red,-{Stealth[length=3mm,width=3mm]},line width=2.5pt]
 (\x,-0.5,0)--(\x,0.5,0);
\foreach \x in {0,...,8}
\draw[fill=black] (\x,0,0) circle (4pt);
\end{tikzpicture}}
\resizebox{0.45\textwidth}{!}{
\begin{tikzpicture}[scale=1.5]
\draw[thick,->] (-1,0,0) -- (9,0,0) node[anchor=west] {$x$};
\draw[thick,->] (0,-1,0) -- (0,1,0) node[anchor=south] {$z$};
\draw[thick,->] (0,0,1) -- (0,0,-1) node[anchor=west] {$y$};
\draw[color=red,-{Stealth[length=3mm,width=3mm]},line width=2.5pt]
 (-0.5,0,0)--(0.5,0,0);
\draw[color=red,-{Stealth[length=3mm,width=3mm]},line width=2.5pt]
 (1,0,-0.75)--(1,0,0.75);
\draw[color=red,-{Stealth[length=3mm,width=3mm]},line width=2.5pt]
 (2.5,0,0)--(2-0.5,0,0);
\draw[color=red,-{Stealth[length=3mm,width=3mm]},line width=2.5pt]
 (3,0,0.75)--(3,0,-0.75);
\draw[color=red,-{Stealth[length=3mm,width=3mm]},line width=2.5pt]
(3.5,0,0)--(4.5,0,0);
\draw[color=red,-{Stealth[length=3mm,width=3mm]},line width=2.5pt]
 (5,0,-0.75)--(5,0,0.75);
\draw[color=red,-{Stealth[length=3mm,width=3mm]},line width=2.5pt]
 (6.5,0,0)--(5.5,0,0);
\draw[color=red,-{Stealth[length=3mm,width=3mm]},line width=2.5pt]
 (7,0,0.75)--(7,0,-0.75);
\draw[color=red,-{Stealth[length=3mm,width=3mm]},line width=2.5pt]
 (7.5,0,0)--(8.5,0,0);
\draw[smooth,dashed,gray,line width=0.75pt]
 plot coordinates {
  (0.5,0,0)
  (1,0,0.75)
  (1.5,0,0)
  (3,0,-0.75)
  (4.5,0,0)
  (5,0,0.75)
  (5.5,0,0)
  (7,0,-0.75)
  (8.5,0,0)
        };
\foreach \x in {0,...,8}
\draw[fill=black] (\x,0,0) circle (4pt);
\end{tikzpicture}}
\caption{\label{fig:vvssz} Visualization of the conventional $S^z$ basis and
  the chiral basis. \textbf{Left panel}: From top to bottom, the states are
  the all spin-down state $|\varnothing\>_z$, the one spin-flip state $|n_1\>_z$, and the all spin-up state $|1,2,\ldots,N\>_z$.
  \textbf{Right panel}: From top to bottom, the spin-helix state
  $|1;\varnothing\>$, the one kink state $|1;n_1\>$, and another spin-helix
  state with different chirality $|1;1,2,\ldots,N\>$.}
\end{figure}

The state $|-1; \nvn\>$ is obtained from  $|1;\nvn\>$ by a simultaneous
inversion of all spins in the $xy$-plane. Such an inversion preserves the
number and the locations of kinks and results in a linearly independent
eigenstate of $V$ with the same eigenvalue. The states with no kinks, in
which the phase winds anticlockwise from every site to the next, play the
role of the ``pseudo vacua'' for the chiral basis. We shall denote them
by $|\k; \varnothing\>$, $\k = \pm 1$. For a visualization of the eigenstates
of $V$ and of $S^z$ see Fig.~\ref{fig:vvssz}.

The operator $V$ counts the number of rotations of the polarization
in the $xy$-plane in a basis state (\ref{cpdstate}). For this reason
we call it the winding number operator and its eigenvalue the
winding number. The XX Hamiltonian $H_{XX}$ commutes with the winding
number operator $V$, if the number of lattice sites $N$ is even.
Hence, on lattices with an even number of sites, $H_{XX}$ can be
diagonalized at fixed winding number. For compatibility with periodic
boundary conditions the winding number must be an integer, implying
that $M$ must have the same parity as $N/2$. Such values of $M$, odd
integers for $N/2$ odd and even integers for $N/2$ even, will be
called admissible.

\begin{theorem}\label{thm2} \cite{2024ChiralBasis}.
For $M=1,2, \ldots,N$ define $\vupb = (p_1, p_2, \ldots, p_M)$, $p_j
\in (-\pi,\pi]$ with $p_1  < \dots < p_M $.
The states
\begin{align}
\begin{aligned}
     |\mu_M(\vupb)\rangle &=
        \sum_{1 \le n_1 < \dots < n_M \le N} \sum_{\kappa = \pm 1}
	\chi_{\nvn} (\vupb)
	\kappa^\xi |\kappa;\nvn\rangle \epc \qd \xi=0,1 \epc \qd
        \chi_{\nvn} (\vupb) &= \frac{1}{\sqrt{2 N^M}} \,
	   \det_{j,\ik = 1, \dots, M} \bigl\{\re^{\i p_j n_\ik}\bigr\} \epc
\end{aligned}
\end{align}
where $M$ is admissible and the chiral quasi-momenta
$p_j$ satisfy $\re^{\i N p_j} = (-1)^{\xi+1}$
for all $p_j$, form a complete set of mutually orthogonal,
normalized eigenstates of the Hamiltonian (\ref{def:XX0}).
The corresponding eigenvalues are
\begin{equation}
     E(\vupb) = 4\JJ \sum_{j=1}^{M} \cos p_j.
\end{equation}
\end{theorem}

We call the basis $\{\mu_M(\vupb)\rangle\}$ the chiral eigenbasis
of the XX Hamiltonian (\ref{def:XX0}). It is of the same appealing
simplicity as the $S^z$ eigenbasis in Thm.~\ref{thm1}.

Unlike the basis of $S^z$ eigenvectors (\ref{szbasis}) (where a
local flip of any spin in vertical direction changes the
$z$-magnetization by one unit), the chiral basis vectors
(\ref{cpdstate}) have a topological nature: a single kink
cannot be added to (removed from) a periodic chain by any local
operatorial action. In an open chain this is only possible at
the boundaries. Thus, the sectors with different chiralities
$M,M\pm1$ are topologically protected.

Note that $[H_{XX}, S^z] = [H_{XX}, V]=0$, but $[V, S^z] \ne 0$.
The two families of XX eigenstates, given by the Theorems \ref{thm1} and
\ref{thm2} are different. This is possible due to the large degeneracy
of the eigenstates of the Hamiltonian. Still, some important eigenstates
are non-degenerate, for instance, the ground state. Within both
eigenbases ($S^z$ and chiral) the ground state
is constructed by filling a Fermi sea with quasiparticles that add
negative contributions to the energy ($\cos \lp_j<0$ in the $S^z$
eigenbasis, or $\cos p_j<0$ in the chiral eigenbasis). It can be
easily seen that for an XX chain of even length $N$, the ground state
belongs to the intersection of the blocks $m=N/2$ of vanishing
magnetization and $M = N/2$ of zero winding number.

\section{\boldmath Comparison of expectation values of observables
in $S^z$ and chiral XX eigenstates}
Here we compare one-point and neighbor two-point functions calculated
in the eigenstates $\ket{\xi_m(\lvp)}$ and $\ket{\mu_M(\vupb)}$. We
shall introduce shorthand notations for the respective expectation values,
\begin{equation}
     \langle \hat A \rangle_{\lvp} \equiv \bra {\xi_m(\lvp)} \hat A \ket{\xi_m(\lvp)}
\end{equation}
in sectors of fixed magnetization and
\begin{equation}
     \langle \hat A \rangle_{\vupb} \equiv \bra {\mu_M(\vupb)} \hat A \ket{\mu_M(\vupb)}
\end{equation}
for fixed winding number. The results are summarized in Tab.~\ref{Table1}.

We are denoting the translation operator by $T$. Note that $[H_{XX},\,T] = 0$,
but $[V,\,T] \ne 0$. We rather have $[V,\,T^2] = 0$. Hence, the usual momentum
is not a good quantum number for chiral eigenstates. We shall further discuss
this below, when we are discussing the elementary excitation of the more
general XY chain.
\begin{table}
\begin{tabular}{|c|c|c|}
\hline
Operator of an observable/conserved quantity$\hat A$  &
$S^z$ eigenstate expectation $\langle \hat A \rangle_{\lvp}$  &Chiral eigenstate
expectation $\langle \hat A \rangle_{\vupb}$  \\[2pt]
\hline 
transversal magnetization & 0   & 0 \\[2pt]
$\sigma_{j_1}^{\alpha_1}\sigma_{j_2}^{\alpha_2}\cdots
\sigma_{j_{2k+1}}^{\alpha_{2k+1}},\,\,\alpha_l=x,y$ & 0 & 0 \\[2pt]
$S^z= \2 \sum_\ik \si_\ik^z$ & $m-N/2$  & $\sum_{j=1}^M \sin p_j$ \\
winding number $V$ &  $\frac12 \sum_{j=1}^m \sin \lp_j$& $\frac{1}{4}(N-2M)$ \\[2pt]
energy & $4\JJ \sum_{j=1}^m \cos \lp_j$   & $4\JJ \sum_{j=1}^M \cos p_j$ \\[2pt]
momentum ($-\ir \ln T$) \  &$\sum_{j=1}^m \lp_j$   & -- \\[2pt]
``double" momentum ($-\ir\ln T^2$) \  &$2\sum_{j=1}^m \lp_j$   & $2\sum_{j=1}^M p_j+(M+\xi) \pi$   \\[2pt]
total magnetization current $\currJ = \sum_\ik j_\ik$  & $8 \sum_{j=1}^m \sin \lp_j$  &$2N \left(1- \frac{4}{N} \sum_{j=1}^{M} \sin^2 p_j  \right)$  \\[2pt]
\hline 
\end{tabular}
\caption{Comparison of observables and conserved quantities
$\langle \hat A \rangle_{\lvp}$  and $\langle \hat A \rangle_{\vupb}$ 
for the $S^z$ and chiral eigenstates of the XX model.}
\label{Table1}
\end{table}

The conserved magnetization current in $z$-direction will be denoted
$\currJ$. Its density is
\begin{equation} \label{jdens}
     j_\ik = 2 \JJ  (\si_\ik^x \si_{\ik+1}^y - \si_\ik^y \si_{\ik+1}^x) \epp
\end{equation}
We can express it as $\currJ = 8\JJ  (V + T V T^{-1})$. The vectors in the $S^z$
eigenbasis are joint eigenvectors of $H_{XX}$, $S^z$ and $T$, implying that
$\<\currJ\>_{\lvp} = 8\JJ  \<(V + T V T^{-1})\>_{\lvp} = 16\JJ  \<V\>_{\lvp}$.

To get a better understanding of the difference of the expectation
values in the two bases, let us consider averages and distributions
of currents within blocks of fixed magnetization and of fixed winding
number.
The average of the total current within any block of fixed
$S^z$ is zero, since the current is an odd function of the symmetrically
distributed quasimomenta $\lp_j$. Indeed, for the XX eigenfunctions
$|\xi_{m}(\lvp)\rangle$ (\ref{XXEigenstatesConventional}) from Theorem~\ref{thm1}
obtained via the usual computational basis, i.e.\ within invariant
blocks of fixed $z$-axis magnetization, $(\sum_{j=1}^N \si_j^z)
\ket{\xi_{m}(\lvp)} = (2m-N) \ket{\xi_{m}(\lvp)}$, $m =0,1,2,\ldots N$,
the average current and its variance within the blocks of fixed $m$ are
\begin{equation}
     {\bar J }_{\mbox{ fixed $m$}}=0, \quad
        \langle(J-\bar J)^2\rangle_m = {32 \,m}\,  \frac {1-\frac{m}{N}}{ 1-\frac{1}{N}}.
 \label{eq:JvarianceExactComputationalBasis}
\end{equation}

\begin{figure}
\centerline{
\includegraphics[width=0.48\textwidth]{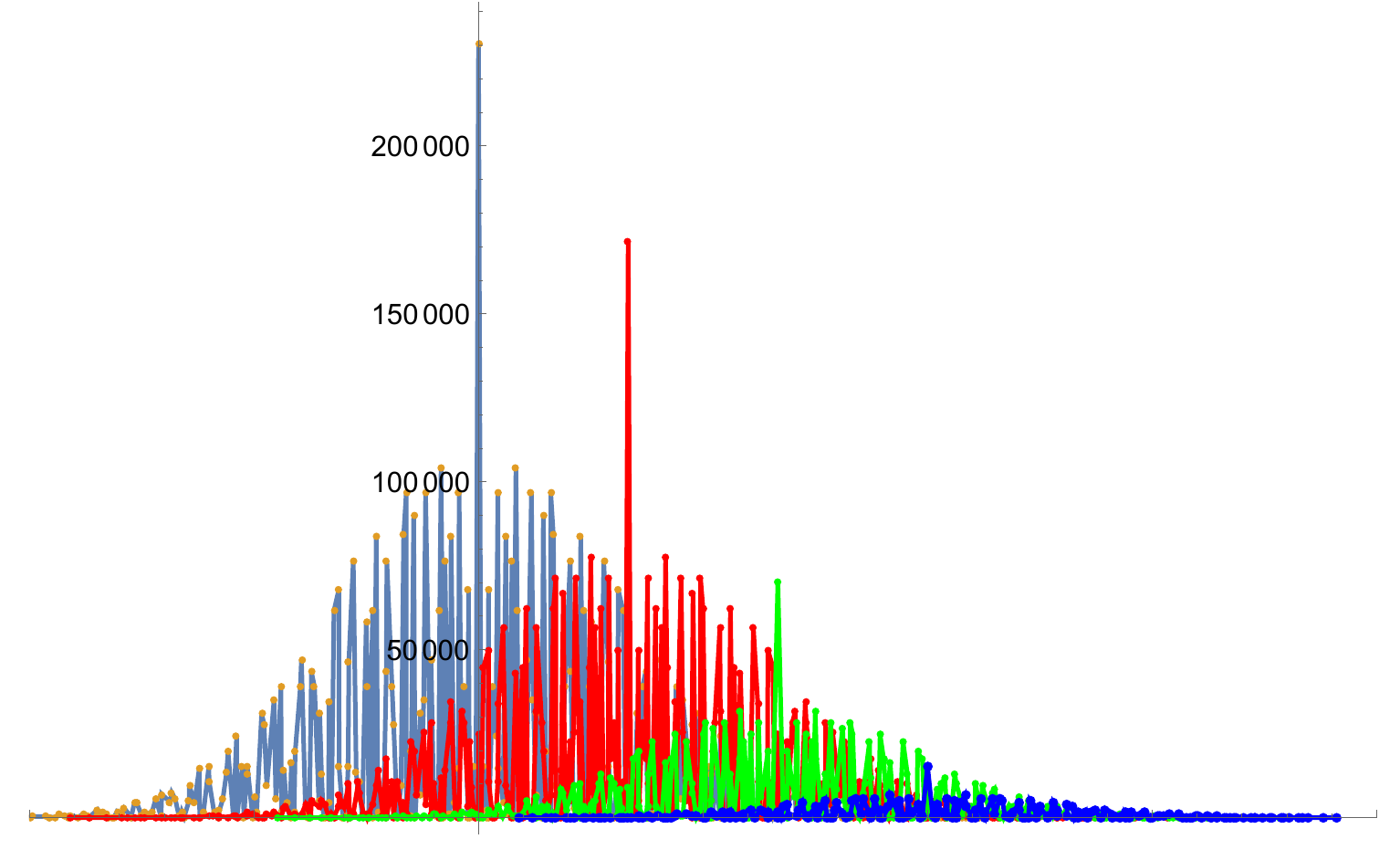}~~~
\includegraphics[width=0.48\textwidth]{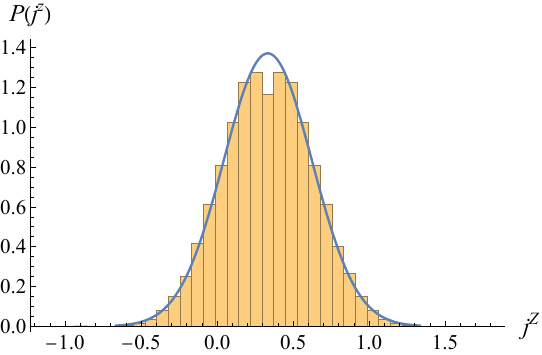}
}
\caption{ \textbf{Left Panel:} Cumulative frequency distribution
(total number of appearances) of the $z$-magnetization current of the 
XX eigenstates inside chiral blocks for a system of size $N=24$. 
Light blue, red, green, blue peaks belong to chiral blocks $M=12,10,8,6$
respectively. The currents' arrangement is symmetric with respect
to the central peak,  located at $\langle j_\ik \rangle = 2(1-\frac{2M}{N})$,
according to Eq.~(\ref{eq:AvCurrent}). The currents for $M > N/2$ (not
shown) are located symmetrically on the left side. 
\textbf{Right Panel:} Probability density distribution of the currents
in the red block of the left Panel. A Gaussian blue curve with exact
variance $\si=\sqrt{35/414} \approx 0.29$ calculated from
(\ref{eq:JvarianceExact}) is a guide to the eye.}
\label{FigCurrDistribution}
\end{figure}

On the contrary, the same average of the total current within chiral
blocks for $\ket{\mu_M(\vupb)}$ is generically nonzero (see App.~\ref{App;Current})
as here the current is the sum of an even function of the quasimomenta $p_j$.
\begin{equation}
{\bar\currJ }=2\JJ  \left(N-{2M} \right)
\label{eq:AvCurrent}
\end{equation}
with variance
\begin{equation}
\langle(J-\bar J)^2\rangle_M = {8M}\,  \frac {1-\frac{M}{N}}{ 1-\frac{1}{N}}.
 \label{eq:JvarianceExact}
\end{equation}
The above result is exact in the thermodynamic limit and numerically appears
to be extremely accurate for all $M, N$.
Examples for the asymmetric distribution of the current for
different fixed winding numbers are shown in Fig.~\ref{FigCurrDistribution}.

\section{Diagonalization of the XY model in the chiral basis}
It is straightforward to see (cf.\ App.~\ref{app:winding}) that the
Hamiltonian
\begin{equation} \label{def:XY}
     H_{XY} = \sum_{n=1}^{N} (J_x \si_n^x \si_{n+1}^x + J_y  \si_n^y \si_{n+1}^y),
              \quad \si_{N+1}^\al \equiv \si_1^\al,
\end{equation}
of the XY spin-$\frac12$ chain with periodic boundary conditions commutes
with the winding number operator $V$, if $N$ is even. Hence, $H_{XY}$ is
block diagonal in the chiral basis. An orthonormal chiral eigenbasis is
obtained by taking the XY limit in the coordinate Bethe Ansatz solution
\cite{2024PedestrianXYZ} of the XYZ chain that was recently obtained by
three of the authors (cf.~App.~\ref{app:reviewXYZchiral}).

The description of the chiral eigenbasis in terms of rapidity
variables below involves the Jacobi theta functions $\theta_j (u,q)$,
$j = 1, 2, 3, 4$, \cite{WatsonBook}. We recall their definition in
App.~\ref{app;Theta} (for more details see also \cite{Mumford82}).
For the nome $q$ we employ the parameterization $q = \eE^{\ir\pi\tau}$,
with $\tau \in {\mathbb C}$, $\Im\tau>0$. We shall further use the
the shorthand notations
\begin{equation}
     \theta_j (u)\equiv\theta_j (u,\eE^{\ir\pi\tau}) \epc \quad
     \theta_j' (u)\equiv\frac{d \theta_j (u)}{d u} \epc \quad
     j=1,2,3,4 \epp
\end{equation}
Restricting ourselves to real positive values of the exchange constants
$J_x$ and $J_y$ in (\ref{def:XY}) for simplicity, we adopt the
parametrization
\begin{equation} \label{eq:JxJyParametrization}
     J_x = \frac{\ell{3}(0)}{\ell{4}(0)} \epc \qd
     J_y = \frac{\ell{4}(0)}{\ell{3}(0)} \epp
\end{equation}
Then the following theorem holds.
\begin{theorem}\label{thm3}
The states
\begin{align} \label{BetheState}
\begin{aligned}
     |\Psi_M(\vvu)\rangle & =
        \sum_{1 \le n_1 < \dots < n_M \le N} \sum_{\kappa = \pm 1}
	   \chi_{\mathbf{n}}(\vvu) \, \kappa^\xi \ket{\kappa;\nn} \epc
	   \qd \xi=0,1 \epc \\
     \chi_{\nn}(\vvu) &= \frac{1}{\sqrt{2N^M}}
        \det_{j,\ik = 1, \dots, M} \bigl\{ \A_{n_j}(\vu_{\ik})\bigr\} \epc \qd
     \A_n(\vu) = \left[\frac{\ell{1}(\vu+\frac{1}{4})}{\ell{1}(\vu-\frac{1}{4})}\right]^{n}
                 \left[\frac{\ell{3}(\vu+\tfrac{\eps_n}{4})}
	                    {\ell{3}(\vu-\tfrac{\eps_n}{4})}\right]^{\frac12},
			    \quad \eps_n=(-1)^n \epc
\end{aligned}
\end{align}
where $M$ is admissible and the ordered Bethe roots $\vvu =
(\vu_1, \vu_2, \ldots,\vu_M)$ satisfy
\begin{equation} \label{baxyelliptic}
     \left[\frac{\ell{1}(\vu_j+\frac{1}{4})}{\ell{1}(\vu_j-\frac{1}{4})}\right]^N=(-1)^{\xi+1},\quad j=1,\ldots,M,
\end{equation}
form a complete set of mutually orthogonal, normalized
eigenstates of the Hamiltonian (\ref{def:XY}), (\ref{eq:JxJyParametrization}).

The corresponding eigenvalues of $H_{XY}$  are 
\begin{equation} \label{eq:XYenergy}
     E(\vvu) = 2 \frac{\ell{2}(0)}{\ell{1}'(0)}
                \sum_{j=1}^M\left[\frac{\ell{1}'(\vu_j-\frac14)}{\ell{1}(\vu_j-\frac14)}
		-\frac{\ell{1}'(\vu_j+\frac14)}{\ell{1}(\vu_j+\frac14)}\right].
\end{equation}
\end{theorem}
\begin{rem}
We consider the Hermitian case, where $\tau$ is purely imaginary.
Without loss of generality, all Bethe roots are restricted to
the rectangles $-\frac12< \Re \vu\leq 0,\,\, -\frac{\Im \tau }{2}<
\Im \vu\leq\frac{\Im \tau}{2}$ and $0< \Re \vu\leq \frac12,\,\,
-\frac{\Im \tau}{2}\leq \Im \vu<\frac{\Im \tau}{2}$.
In our elliptic parametrization the Bethe roots $\{\vu_j\}$ are then
distributed along two lines in the complex plane: $\Re(\vu_j)=0,\frac12$
(cf.~App.~\ref{app:goelliptic}).
\end{rem}

\begin{rem}
Once $\vu_j$ is a solution of the Bethe Ansatz equations
(\ref{baxyelliptic}), then $\vu_j+\frac12$ is a solution as
well. Since $\vu_j$ and $\vu_j+\frac12$ yield energies of the same
absolute values, but with different signs, the complete spectrum of
the Hamiltonian is symmetric with respect to sign reversal.
\end{rem}

\begin{rem}
As in the XY case, our chiral Bethe states are eigenstates of the
Hamiltonian but not of the transfer matrix of the underlying eight-vertex
model. This distinguishes our approach from
the algebraic Bethe Ansatz method \cite{TaFa79,2020SlavnovXY}. 
\end{rem}

The proof of Thm.~\ref{thm3} is split into three parts: the proof
of the commutativity $[H_{XY},V]=0$, the proof of the main body of the
theorem for unnormalized eigenvectors, and, finally, the proof of
the normalization. The three proofs are given in App.~\ref{app:xylimitofxyz}.

The form (\ref{baxyelliptic}) of the Bethe equations suggests to
introduce a ``momentum function'' $\pu$ by the equation
\begin{equation} \label{momfun}
     \re^{\i \pu (\vu)} = \frac{\th_1 \bigl(\vu + \4\bigr)}{\th_1 \bigl(\vu - \4\bigr)}
\end{equation}
and to express the eigenvectors and the energy eigenvalues in
terms of the quasi-momenta $\vup_j = \pu (\vu_j)$. This is done
in App.~\ref{app:goelliptic} and results in the following corollary
to Thm.~\ref{thm3}.

\begin{cor}\label{thm4}
The states
\begin{align} \label{BetheState;2}
\begin{aligned}
     |\widetilde\Psi_M(\vupb)\rangle & =
        \sum_{1 \le n_1 < \dots < n_M \le N} \sum_{\kappa = \pm 1}
	   \widetilde\chi_{\mathbf{n}}(\vupb)\,
	   \kappa^\xi  \ket{\kappa;\nn} \epc \qd \xi=0,1,\qd
     \widetilde\chi_{\nn}(\vupb) = 
        \frac{1}{\sqrt{2N^M}}\,\det_{j,\ik = 1, \dots, M}
	   \left\{ \widetilde{\A}_{n_j}(\vup_{\ik})\right\},\\
     \widetilde{\A}_n(\vup) & =
        \exp[\ir(n\vup-\eps_n\varphi(\vup))] \epc \quad 
     \varphi(\vup)=\frac12\arctan\left(\frac{J_x-J_y}{J_x+J_y}\tan \vup\right),\quad \eps_n = (-1)^n \epc
\end{aligned}
\end{align}
where $M$ is admissible and the quasi-momenta $\vupb =
(\vup_1,\vup_2,\ldots,\vup_M)$ satisfy $\eE^{\ir N\vup_j}=(-1)^{\xi+1}$ for all $\vup_j$, form a complete
set of mutually orthogonal, normalized eigenstates of the Hamiltonian
(\ref{def:XY}), (\ref{eq:JxJyParametrization}). When rewritten in terms
of the quasi-momenta $\{\vup_j\}$ the energy takes the form
\begin{equation} \label{energy;XY}
     \widetilde{E}(\vupb) = \sum_{j=1}^M\pm2\sqrt{4\cos^2\vup_j+\left({J_x-J_y}\right)^2} \,,
\end{equation}
where $+$ holds for $\vup_j\in(-\pi/2,\pi/2]$ and $-$ for
$\vup_j\in(\pi/2,3\pi/2]$. 
\end{cor}

\begin{rem}
At the discontinuous points $p=\frac{\pi}{2},\frac{3\pi}{2}$, the function 
$\varphi$ takes the specific values $\varphi(\frac{\pi}{2})=
\varphi(\frac{3\pi}{2})=\frac{\pi}{4}$.
\end{rem}

\begin{rem}
Corollary \ref{thm4} and Theorem \ref{thm2} are
strikingly similar, e.g., the Bethe Ansatz equations for the different sets $\{p_j\}$
are of the same form. As $J_{x,y}\to1$
we have 
$\widetilde{\A}_n(p_j) \rightarrow \eE^{\ir n\vup_j}$ (for most cases), so the
XY eigenvectors reduce to the XX eigenvectors. More details are shown
in App.~\ref{app;XX-limit}. 
\end{rem}

\begin{figure}
\begin{center}
\begin{tabular}{ll}\includegraphics[width=.47\textwidth]{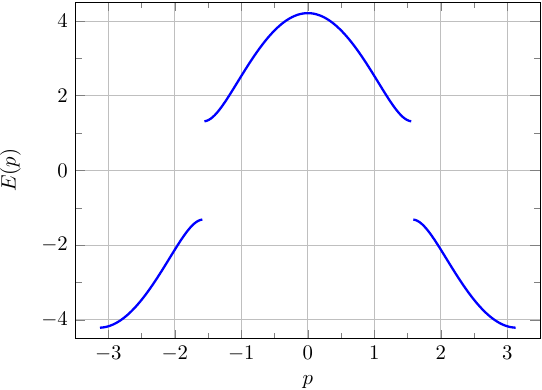}
&~~~ \includegraphics[width=.47\textwidth]{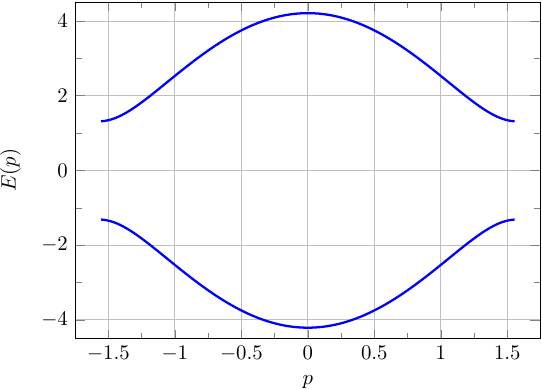}
\end{tabular}
\caption{\label{Fig:disprels} The dispersion relations in the full
Brillouin zone $(-\p,\p]$ (left panel) and in the reduced Brillouin
zone $(- \frac{\pi}{2}, \frac{\pi}{2}]$ (right panel) for $J_x=J_y^{-1}
= 1.38$. Note that each subband shows $\pi$-periodicity. Here
$E(p) = \pm 2 \sqrt{4 \cos^2 (p) + (J_x - J_y)^2}$, where the plus
sign refers to the upper branch in the figure, while the minus sign
refers to the lower branch.}
\end{center}
\end{figure}

The dispersion relation of our elementary excitations, shown in the
left panel of Fig.~\ref{Fig:disprels}, does not look like that 
of a single particle. It resembles the dispersion relation of
phonons in a harmonic chain in which every second mass has been
altered, implying that the size of the unit cell is doubled, while the
size of the Brillouin zone is reduced to half of its value. A band
gap then opens in the reduced zone scheme and the dispersion is split in
an acoustic and an optical branch, representing two different species
of phonons, optical and acoustic.

In our case a reduction of the energy-momentum dispersion for all momenta in the
interval $(-\pi,+\pi]$ to two subbands in $(-\pi/2,+\pi/2]$ is fully consistent with the chiral Bethe
Ansatz solution: the momentum $\pu$ with values in
$(-\pi/2,+\pi/2]$ or $(+\pi/2,+3\pi/2]$ is well-defined, but only $2\pu$ has a physical
meaning. The logarithmic form of the Bethe equations reads for $\xi=0$ resp.~$\xi=1$
\begin{equation}
     \pu_l = 2n_\ik \frac{\p}{N} \mod 2\p \epc \qd \text{resp.} \qd
     \pu_l= (2n_\ik - 1) \frac{\p}{N} \mod 2\p \qd
     \text{for $n_\ik \in {\mathbb Z}$,\quad $l= 1, \dots, M$.}
\end{equation}
Since $N$ is even,
$p_l$ and $p_l+\pi$ belong to the same sector (the same value
of $\x$). This establishes a one-to-one correspondence between the
quasi-momenta $p_l$ and $p_l+\pi$ ($-\pi/2<p_l\leq \pi/2$). The corresponding wave functions are related by
\begin{equation}
     \widetilde{\cal A}_n \bigl(p_l+\pi) =
        \re^{\i [n ( p_l+\p) - \eps_n \ph( \pu_l+\p)]} =
        (-1)^n \re^{ \i [n \pu_l - \eps_n \ph(\pu_l)]} =
        (-1)^n \widetilde{\cal A}_n(\pu_l),\quad  -\pi/2<p_l\leq\pi/2\,,
\end{equation}
in full accord with (\ref{amplstatewithr}).
We further recall that $p_l$ and $p_l+\pi$ yield energies with opposite signs. This means that the particle content of
any state can be characterized by two numbers, a momentum
$\pu_l \in (- \frac{\p}{2}, \frac{\p}{2}]$ and a ``charge''
$c_\ik \in \{-1, 1\}$. The wave functions are
\begin{equation}
     \widetilde{\cal A}_n \bigl(\vup_\ik, c_\ik) =
        c_\ik^n \re^{\i [n \vup_\ik- \eps_n \ph(\vup_\ik)]}\,,
\end{equation}
and the corresponding energies
\begin{equation}
E=2\sum_{l=1}^Mc_l\sqrt{4\cos^2\vup_l+\left({J_x-J_y}\right)^2} \epp
\end{equation}
Note that this interpretation is well compatible with the
fact that the basis states and the vacuum we are building
on are invariant under a translation by \textit{two} lattice
sites. Since excitations of the Hamiltonian consist of pairs
of kinks, their momentum still ranges from $- \p$ to $\p$.
Our kinks are very much like the spinons occurring in the
massive XXZ chain in the thermodynamic limit \cite{JiMi95}.
They can only be created in pairs, the momentum of a single
spinon ranges from $- \frac{\p}{2}$ to $\frac{\p}{2}$, and they
live over a pair of degenerate vacua given by two appropriately
deformed N\'eel states. In our case the occurance of such
excitations is not restricted to the thermodynamic limit.
The kinks do exist as bare particles on finite chains, which
we find quite interesting.

\section{XY chiral eigenfunctions: observables}
In the XY case we are as well able to calculate some simple expectation
values in chiral eigenstates, which again can be presented in terms
of sums of functions of quasimomenta $\vup_j$. In Tab.~\ref{Table2} we
compare some observables and conserved quantities for the XX and XY model.
A crucial difference between the two models consists in the fact that
$S^z$ is not conserved in the XY model. As a consequence, $j_\ik$ as
defined in (\ref{jdens}) cannot be interpreted as magnetic current
density, but is just a specific two-point operator.
Proofs of the statements in the right column of Tab.~\ref{Table2} are
given in Apps.~\ref{app;one-point} and \ref{app;two-points}. When one
of the periods of the theta functions $\tau\to+\ir\infty$, the XY model
reduces to the XX model with $J_{x,y}\to 1$.

\begin{table}[htbp]
\begin{tabular}{|c|c|c|}
\hline
Operator of an observable/conserved quantity$\hat A$  &  XX model expectation $\langle \hat A \rangle_{\vupb}$ & XY model expectation $\langle \hat A \rangle_{\vupb}$ \\[2pt]
\hline 
transversal magnetization & 0 & 0  \\[2pt]
energy & $4\sum_{j=1}^M \cos p_j$ & $4\sum_{j=1}^M\pm\sqrt{\cos^2\vup_j+\left(\frac{J_x-J_y}2\right)^2}$  \\[2pt]
``double" momentum ($-\ir \ln T^2$) \  & $2\sum_{j=1}^M p_j+(M+\xi)\pi$   & $2\sum_{j=1}^M \vup_j+(M+\xi)\pi$\\[2pt]
$ \si_{n}^z$ & $\frac{2}{N}\sum_{j=1}^M \sin p_j$ & $\frac{J_y^{\epsilon_n}}{N}\sum_{j=1}^M\pm{\sin (2\vup_j) }/\sqrt{\cos^2\vup_j+\left(\frac{J_x-J_y}2\right)^2}$  \\[2pt]
winding number $V$ & $\frac{1}{4}(N-2M)$ &  $\frac{1}{4}(N-2M)$\\[2pt]
$\si_\ik^x \si_{\ik+1}^y - \si_\ik^y \si_{\ik+1}^x$  &$ 1- \frac{4}{N}\sum_{j=1}^{M} \sin^2 p_j $  & $1- \frac{4}{N} \sum_{j=1}^{M} \sin^2\vup_j $\\[2pt]
\hline 
\end{tabular}

\caption{Comparison of observables and conserved quantities for chiral XX and
  XY eigenstates. Here the sign $+$ holds for $\vup_j\in(-\pi/2,\pi/2]$ and
  $-$ for $\vup_j\in(\pi/2,3\pi/2]$.}
\label{Table2}
\end{table}

\section{Discussion}
We have constructed a chiral eigenbasis for the XY quantum spin chain,
i.e., an orthonormal basis of eigenstates of the XY Hamiltonian that
are classified according to their winding number. In the limit of
isotropic exchange in the $xy$-plane we recover the chiral eigenbasis
previously constructed \cite{2024ChiralBasis} for the XX chain. Our
solution does not, at any stage, rely on a mapping to non-interacting
Fermions. It rather is of Bethe Ansatz type and is a limit of an
alternative Bethe Ansatz solution of the more general XYZ quantum
spin chain, recently obtained by three of the authors in 
\cite{2024PedestrianXYZ}.

We like to interpret the Bethe eigenstates constructed in
\cite{2024PedestrianXYZ} and reviewed in App.~\ref{app:reviewXYZchiral}
as a chiral eigenbasis for the XYZ model. What is missing so far to
complete this interpretation is the construction of an appropriate
winding number operator, which remains as a challange for future work.
The very fact that such an alternative Bethe Ansatz exists is
of some interest in view the difficulties that seem to persist in attempts
(see e.g.\ \cite{2020SlavnovXY,2023SlavnovXY}) to use
the previous Bethe Ansatz solutions \cite{TaFa79,FeVa96} for the
calculation of matrix elements of local operators between eigenstates.
In fact, even the calculation of the norm of the Bethe eigenstates
has remained a problem, even in the XY limit \cite{2023SlavnovXY}.
This problem seems absent in the chiral Bethe Ansatz advocated here
(cf.\ App.~\ref{app:chiralnorm}). It would therefore be interesting
to lift our work to an algebraic form of the Bethe Ansatz for
the XY chain and further on to the more general XYZ chain. This would be
an important step towards extending the algebraic Bethe Ansatz approach
to the calculation of correlation functions, that was successful
for the XXZ chain \cite{KMT99a,KMT99b,GKS04a}, to the XYZ case.

In Ref. \cite{2024ChiralBasis} we have seen, in the special case of the
XX chain, that overlaps needed for the calculation of certain transverse
correlation functions take a simpler form in the chiral eigenbasis
than in the $S^z$ eigenbasis. This allowed us to obtain an
exact and analytic expression for the time evolution of a
one-point function that measures the experimentally realized decay
of spin helices \cite{SHS-Ketterle}. In the present work we have
continued our comparison of the description of observables in the chiral
eigenbasis and in the $S^z$ eigenbasis, mostly by caculating simple
one-point and neighbor two-point correlations in eigenstates of the
respective basis.

The chiral Bethe Ansatz for the XY chain offered an interesting
physical picture. The excitations that make up the chiral eigenbasis
can be interpreted as scattering states of kinks and anti-kinks,
topological excitations that can only by created in pairs (two
kinks, two anti-kinks, or a kink and an anti-kink). Each individual
kink or anti-kink changes the winding number by $\pm \2$, each pair
by $0$ or $1$. The momenta of the individual kinks and anti-kinks
take values in a reduced Brillouin zone $(-\frac{\p}{2},
\frac{\p}{2}]$. The kinks and anti-kinks each carry a charge $+ 1$ or $- 1$.
Depending on this value, their energy is positive or negative. This
is in analogy with the spinons in the XXZ chain \cite{JiMi95} (for
a recent extensive discussion cf.~\cite{Rutkevich22,Rutkevich24})
in its massive antiferromagnetic regime which are generated over
two degenerate N\'eel states of the infinite chain. By way of contrast,
our kinks exist on finite chains.

Being an orthonormal basis with factorized structure, the chiral basis
should be also well suited for numerical calculations. It can be
implemented with usual binary code registers in order to diagonalize any
Hamiltonian, which need not be integrable or even a one-dimensional.
The chiral basis should perform particularly well, when dealing with
problems where topology plays a role, e.g.\ when the initial quantum
state has a helical structure. In this case, as argued in \cite{ChiralMoreStable},
the topological structure gets protected against perturbations like
strong local noise, a phenomenon called helical protection by the
authors of \cite{ChiralMoreStable}. The helical protection entails
that the temporal dynamics involving transitions between chiral sectors
with different winding numbers should be slower than the dynamics
within a chiral sector (i.e.\ not involving transitions between
states of different winding number), at least for small times. For a
future clarification further studies would be necessary, which is, however,
beyond the scope of the present communication.

\section*{Acknowledgment} 
X. Z. acknowledges financial support from the National Natural Science
Foundation of China (No. 12204519). F.~G.\ would like to thank J.-M. Maillet
for a stimulating discussion. 
F. G. and A. K. acknowledge financial support by Deutsche Forschungsgemeinschaft
through FOR 2316. A. K. also gratefully acknowledges support through
a fellowship from the Chinese Academy of Sciences and the Innovation
Academy for Precision Measurement Science and Technology, Wuhan.
V. P. acknowledges support by Deutsche Forschungsgemeinschaft
through DFG project KL 645/20-2 and by an ERC Advanced grant
No.~101096208 -- QUEST.

\bibliographystyle{amsplain}

\bibliography{ChiralBasis}

\appendix

\section{Theta functions}\label{app;Theta}
Following Ref.~\cite{WatsonBook}, but rescaling the arguments
by $\p$ we define the four Jacobi theta functions
\begin{align}
\label{deftheta}
\begin{aligned}
&\theta_{1}(u,q)=2\sum_{n=0}^\infty(-1)^n q^{(n+\frac12)^2}\sin[(2n+1)\p u],\\
&\theta_{2}(u,q)=2\sum_{n=0}^\infty q^{(n+\frac12)^2}\cos[(2n+1)\p u],\\
&\theta_{3}(u,q)=1+2\sum_{n=1}^\infty q^{n^2}\cos(2n\p u),\\
&\theta_{4}(u,q)=1+2\sum_{n=1}^\infty (-1)^nq^{n^2}\cos(2n\p u).
\end{aligned}
\end{align}

\section{\boldmath Commutativity of $H_{XY}$ and $V$}
\label{app:winding}
Here we show that the XY Hamiltonian commutes with $V$ defined in (\ref{eq:Voper}).
Let us split the XY Hamiltonian into two parts, the $X$-part and the $Y$-part,
\begin{align}
H_{XY} = J_x \sum_{n=1}^{N}   \si_n^x \si_{n+1}^x + J_y  \sum_{n=1}^{N} \si_n^y \si_{n+1}^y,
\label{app:XY}
\end{align}
where $N$ is even, and prove the commutativity of each part with
$V$ individually.

For the $X$-part we have
\begin{align}
&4\left[ \sum_{n=1}^{N}   \si_n^x \si_{n+1}^x ,  \,V \right] = 
\left[\sum_{n=1}^{N}   \si_n^x \si_{n+1}^x, \,\,\sum_{\ik=1}^{N/2} \left(  \si_{2\ik-1}^x   \si_{2\ik}^y
- \si_{2\ik}^y   \si_{2\ik+1}^x \right)\right]\notag \\[1ex]
&\qd =\left[ \sum_{n=1}^{N/2}   (\si_{2n-1}^x \si_{2n}^x +  \si_{2n}^x \si_{2n+1}^x), \,\  \sum_{\ik=1}^{N/2} \left(  \si_{2\ik-1}^x   \si_{2\ik}^y
- \si_{2\ik}^y   \si_{2\ik+1}^x \right)\right]\notag \\[1ex]
&\qd = (2\i-2\i)\sum_{n=1}^{N/2} \si_{2n}^z
+ (2 \i-2\i) \sum_{n=1}^{N/2} \si_{2n-1}^x \si_{2n}^z \si_{2n+1}^x=0 \epp
\end{align}
Analogously, we check the commutativity of the $Y$-part in (\ref{app:XY}) with $V$.
Thus, we obtain
\begin{align}
[H_{XY},V] =0, \label{eq:comm0}
\end{align}
and, consequently, the XY-model can be diagonalized within blocks with
a fixed (and admissible) number of kinks $M$.    

\section{Chiral Bethe vectors of the XY model}
\label{app:xylimitofxyz}
\subsection{Chiral coordinate Bethe Ansatz for the periodic XYZ chain}
\label{app:reviewXYZchiral}
Let us first consider the general periodic XYZ chain
\begin{align} 
&H_{XYZ}=\sum_{n=1}^N\left(J_x\,\mathbf{\sigma}_n^x\sigma_{n+1}^x+J_y\,\sigma_n^y\sigma_{n+1}^y+J_z\,\sigma_n^z\sigma_{n+1}^z\right).\label{XYZ}
\end{align}
The exchange
coefficients $\{J_x,\,J_y,\,J_z\}$ are parameterized by the crossing parameter
$\eta$ as \cite{Baxter-book,Wang-book,Cao2013off}
\begin{align}
&J_x=\frac{\ell{4}(\eta)}{\ell{4}(0)},\quad J_y=\frac{\ell{3}(\eta)}{\ell{3}(0)},\quad J_z=\frac{\ell{2}(\eta)}{\ell{2}(0)}.
\end{align}

In Ref. \cite{2024PedestrianXYZ}, the
periodic XYZ chains with $\eta$ satisfying the following root-of-unity condition has been studied
\begin{align}
&(N-2M)\eta =2L \tau + 2K,\quad 0\leq M\leq N,\quad L,K\in\mathbb{Z},\no\\
&s\eta=2 L_0 \tau + 2 K_0,\quad s\in\mathbb{N}^+,\quad L_0,K_0\in\mathbb{Z},\label{RootUnity}
\end{align}
where $s$ is the smallest positive integer satisfying
Eq.~(\ref{RootUnity}).

Define the following local state
\begin{align}
\phi(v)&=\binom{\ell{1}(v,q^2)}{\ell{4}(v,q^2)},\qquad v\in\mathbb{C},\label{psi}
\end{align}
Then we can construct a set of factorized states \cite{2024PedestrianXYZ}
\begin{align}
&|d;n_1,n_2,\ldots,n_M\rangle_{xyz}\no\\
=&\bigotimes_{\ik_1=1}^{n_1}\phi(v_{2d+\ik_1})\bigotimes_{\ik_2=n_1+1}^{n_2}\phi(v_{2d+\ik_2-2})\cdots \no\\
&\cdots\bigotimes_{\ik_M=n_{M-1}+1}^{n_M}\phi(v_{2d+\ik_M-2M+2})\bigotimes_{\ik_{M+1}=n_{M}+1}^{N}\phi(v_{2d+\ik_{M+1}-2M}),\label{eq:Basis}\\[4pt]
&1\leq n_1<n_2<\ldots<n_M\leq N.\no
\end{align}
where $v_m=v_1+(m-1)\eta$.
Under the root-of-unity condition (\ref{RootUnity}), the vectors
\begin{align}
&\ket{d;n_1,n_2,\ldots,n_M}_{xyz},\quad d=0,1,\ldots,s-1,\no\\
&1\leq n_1< n_2<\ldots<n_M\leq N,\label{Basis;XYZ}
\end{align} form an invariant subspace of the Hilbert space.
The eigenstates of $H_{XYZ}$ thus can be  expanded as 
\begin{align}
&|\Psi_M(\vvu)\rangle_{xyz}=
\sum_{d=0}^{s-1}\,\sum_{\nn}F_{M,d,\nn}(\vvu)\ket{d;n_1,\ldots,n_M}_{xyz},\label{BS;XYZ}
\end{align}
where $\nn=\{n_1,\ldots,n_M\}$ with $1\leq n_1<n_2\cdots<n_M\leq N$, which applies to the following sections in the appendix. The expansion coefficients $\{F_{M,d,\nn}(\vvu)\}$ are given by the chiral coordinate Bethe Ansatz method \cite{2024PedestrianXYZ}. In this paper we only consider the case where $\eta$ is real ($L = L_0 = 0$ in Eq.~\eqref{RootUnity}), and $\{F_{M,d,\nn}(\vvu)\}$ now reads
\begin{align}
&F_{M,d,\nn}(\vvu)=\eE^{-d\omega}\sum_{x_1,\ldots,x_M}C_{x_1,\ldots,x_M}\prod_{\ik=1}^{M} U_{2d-2\ik+n_\ik+2}^{(n_\ik)}(\vu_{x_\ik}),
\label{M;Ansatz}
\end{align}
where $U_{m}^{(n)}(v)$ is defined by
\begin{align}
U_{m}^{(n)}(v)=\left[\frac{\ell{1}(v+\frac{\eta}{2})}{\ell{1}(v-\frac{\eta}{2})}\right]^{n}\frac{\ell{2}(v-v_{m}+\frac{\eta}{2})}{\ell{2}(v_{m-1})\ell{2}(v_{m})},\label{def;U}
\end{align} 
and $\{x_1,\ldots,x_M\}$
is a permutation of $\{1,\ldots,M\}$. The coefficients
$\{C_{x_1,\ldots,x_M}\}$ in terms of Bethe roots $\{\vu_1,\ldots,\vu_M\}$
satisfy
\begin{align}
\frac{C_{\ldots,x_{n+1},x_n,\ldots}}{C_{\ldots,x_{n},x_{n+1},\ldots}}&=\frac{\ell{1}(\vu_{x_n}-\vu_{x_{n+1}}-\eta)}{\ell{1}(\vu_{x_n}-\vu_{x_{n+1}}+\eta)}.
\end{align}
The Bethe roots $\{\vu_1,\ldots,\vu_M\}$ and the parameter $\omega$ satisfy the following Bethe Ansatz equations 
\begin{align}	
&\eE^{\omega} \left[\frac{\ell{1}(\vu_j+\frac{\eta}{2})}{\ell{1}(\vu_j-\frac{\eta}{2})}\right]^N\prod_{\ik\neq j}^M\frac{\ell{1}(\vu_j-\vu_\ik-\eta)}{\ell{1}(\vu_j-\vu_\ik+\eta)}=1,\quad \eE^{s\omega}=1,\quad  j=1,2,\ldots,M.\label{BAE;XYZ}
\end{align}
The energy in terms of the Bethe roots reads 

\begin{align}
E_{\rm{xyz}}(\vvu)=2\frac{\ell{1}(\eta)}{\ell{1}'(0)}\sum_{j=1}^M\left[\frac{\ell{1}'(\vu_j-\frac{\eta}{2})}{\ell{1}(\vu_j-\frac{\eta}{2})}-\frac{\ell{1}'(\vu_j+\frac{\eta}{2})}{\ell{1}(\vu_j+\frac{\eta}{2})}\right]+N\frac{\ell{1}'(\eta)}{\ell{1}'(0)}.\label{energy;XYZ}
\end{align}

\subsection{Reduction from the XYZ Model to the XY Model}
When $\eta=\frac12$, the XYZ chain (\ref{XYZ}) reduces to the XY chain as follows 
\begin{align}
\left\{J_x,\,J_y,\,J_z\right\}=\left\{\frac{\ell{3}(0)}{\ell{4}(0)},\,\frac{\ell{4}(0)}{\ell{3}(0)},\,0\right\}.\label{XY;1}
\end{align}
Consider a periodic XY chain with an even number of sites. Then we
can get the value of $M$ and $s$ in Eq.~(\ref{RootUnity}),
\begin{align}
&s=2,\,\, M=0,2,\ldots,N,\quad \mbox{for}\,\,N=4m,\quad m\in\mathbb{N}^+,\\
&s=2,\,\, M=1,3,\ldots,N-1,\quad \mbox{for}\,\,N=4m+2,\quad m\in\mathbb{N}.
\end{align}
Therefore, we can directly obtain the chiral Bethe state of the XY chain from the results for the XYZ chain. By letting $v_1=\frac{1 + \tau}{2}$ and eliminating an overall factor in (\ref{Basis;XYZ}) and (\ref{BS;XYZ}), we finally arrive at the Bethe state given by Eq.~(\ref{BetheState}).
\begin{rem}
When $\eta=\frac12$, we see that $J_x/J_y>0$.
By letting $\eta=-\frac12-\tau$, the XYZ chain (\ref{XYZ}) reduces to another XY model with
\begin{align}
\{J_x,\,J_y,\,J_z\}=\eE^{-\ir \pi \tau}\left\{\frac{\ell{3}(0)}{\ell{4}(0)},\,-\frac{\ell{4}(0)}{\ell{3}(0)},\,0\right\}.\label{XY;2}
\end{align}
\end{rem}

\subsection{Normalization and orthogonality of Bethe vectors}
\label{app:chiralnorm}
Since the $|\Ps_M (\vvu)\>$ are simultaneous eigenvectors of
$H_{XY}$ and $V$, the scalar products $\<\Ps_M (\vvu)|\Ps_{M'} (\vvu')\>$
with $M \ne M'$ all vanish.

Using the orthonormality of the chiral basis states
\begin{equation} \label{orthochiral}
     \<\iota; \mv|\k; \nv\> = \de_{\iota, \k} \de_{\mv, \nv}
\end{equation}
and the properties of the determinant, we first of all see that
\begin{align} \label{orthoxyeigen}
     \<\Ps_M (\vvu)|\Ps_M (\vvu')\> & =
        \sum_{\substack{1 \le m_1 < \dots < m_M \le N \\
	      1 \le n_1 < \dots < n_M \le N}}
        \sum_{\iota, \k = \pm 1} \chi_\mv (\vvu)^* \chi_\nv (\vvu')\,
	   \iota^\x \kappa^{- \x'} \<\iota; \mv|\k; \nv\> \notag \\
     & = \sum_{1 \le n_1 < \dots < n_M \le N}
        \sum_{\k = \pm 1} \kappa^{\x - \x'} \chi_\nv (\vvu)^* \chi_\nv (\vvu')
       = 2 \de_{\x, \x'} \sum_{1 \le n_1 < \dots < n_M \le N}
         \chi_\nv (\vvu)^* \chi_\nv (\vvu') \notag \\
     & = \frac{\de_{\x, \x'}}{M! N^M} \sum_{\nv \in \{1, \dots, N\}^M}
         \det_M \bigl\{{\cal A}_{n_j} (u_k)^*\bigr\}
         \det_M \bigl\{{\cal A}_{n_j} (u_k')\bigr\} \notag \\
     & = \frac{\de_{\x, \x'}}{M! N^M} \sum_{\nv \in \{1, \dots, N\}^M}
         \sum_{Q \in \mathfrak{S}^M} \sign (Q) \,
	 {\cal A}_{n_{Q1}} (u_1)^* \dots {\cal A}_{n_{QM}} (u_M)^*
         \det_M \bigl\{{\cal A}_{n_j} (u_k')\bigr\} \notag \\
     & = \frac{\de_{\x, \x'}}{M! N^M} \sum_{\nv \in \{1, \dots, N\}^M}
         \sum_{Q \in \mathfrak{S}^M}
	 {\cal A}_{n_{Q1}} (u_1)^* \dots {\cal A}_{n_{QM}} (u_M)^*
         \det_M \bigl\{{\cal A}_{n_{Qj}} (u_k')\bigr\} \notag \\
     & = \frac{\de_{\x, \x'}}{M! N^M} \sum_{Q \in \mathfrak{S}^M}
         \det_M \biggl\{ \sum_{n=1}^N {\cal A}_n (u_j)^* {\cal A}_n (u_k')\biggr\}
       = \frac{\de_{\x, \x'}}{N^M}
         \det_M \biggl\{ \sum_{n=1}^N {\cal A}_n (u_j)^* {\cal A}_n (u_k')\biggr\} \epp
\end{align}
Here we have used (\ref{orthochiral}) in the second equation, the fact
that $\x, \x' \in \{0, 1\}$ in the third equation, the definition
(\ref{BetheState}) of $\chi_\nv (\vvu)$ in the fourth equation, the
definition of the determinant as a sum of permutations in the fifth
equation, the antisymmetry of the determinant in the sixth equation,
the multi-linearity of the determinant in the seventh equation, and
the fact that the number of elements in the symmetric group is $M!$
in the last equation.

It remains to evaluate the sum inside the determinant on the
right hand side of (\ref{orthoxyeigen}). For this purpose we notice
that
\begin{equation} \label{pinsteadofu}
     \sum_{n=1}^N {\cal A}_n (u_j)^* {\cal A}_n (u_k') =
        \sum_{n=1}^N \widetilde {\cal A}_n (p_j)^* \widetilde {\cal A}_n (p_k') =
	\re^{\i (\ph(p_j) - \ph(p_k'))} \sum_{n=1}^{N/2} \re^{\i 2n (p_k' - p_j)} +
	\re^{\i (\ph(p_k') - \ph(p_j))} \sum_{n=1}^{N/2} \re^{\i (2n - 1) (p_k' - p_j)}
	\epc
\end{equation}
where $p_j = p(u_j)$, $p_k' = p(u_k')$ and where we have used
(\ref{BetheState;2}). As a direct consequence of the geometric
sum formula we have the identities
\begin{equation}
     \sum_{n=1}^{N/2} x^{2n} =
        \begin{cases}
	     x^2 \, \frac{1 - x^N}{1 - x^2} & \text{if $x^2 \ne 1$} \\[1ex]
	     \frac{N}{2} & \text{if $x^2 = 1$}
	\end{cases} \epc \qqd
     \sum_{n=1}^{N/2} x^{2n-1} =
        \begin{cases}
	     x \, \frac{1 - x^N}{1 - x^2} & \text{if $x^2 \ne 1$} \\[1ex]
	     x \frac{N}{2} & \text{if $x^2 = 1$}
	\end{cases} \epp
\end{equation}
We shall use these identities for $x = \re^{\i(p_k' - p_j)}$. The
Bethe Ansatz equations in Corollary~\ref{thm4} for $\x = \x'$ then
imply that $x^N = 1$, and
\begin{equation} \label{xsqnone}
     \sum_{n=1}^N \widetilde {\cal A}_n (p_j)^* \widetilde {\cal A}_n (p_k') = 0 \epc
\end{equation}
if $x^2 \ne 1$.

If $x = -1$, then $p_k' - p_j = \p \mod 2\p$; if $x = 1$, then $p_k' - p_j = 0
\mod 2\p$. It follows in both cases from the $\p$-periodicity of the function
$\ph$ that $\ph(p_k') = \ph(p_j)$. Thus,
\begin{equation} \label{xsqone}
     \sum_{n=1}^N \widetilde {\cal A}_n (p_j)^* \widetilde {\cal A}_n (p_k')
        = (1 + x) \frac{N}{2} \epc
\end{equation}
if $x^2 = 1$.

Using now (\ref{pinsteadofu}), (\ref{xsqnone}) and (\ref{xsqone}) in
(\ref{orthoxyeigen}) we have proved the orthonormality condition
$\<\Ps_M (\vvu)|\Ps_M (\vvu')\> = \de_{\vvu, \vvu'}$. Then a simple
counting argument as employed in the XX case in \cite{2024ChiralBasis}
implies that the chiral eigenstates form a basis.

\section{Quasi-momentum representation of Bethe states and dispersion relation for 
the XY model}
\label{app:goelliptic}

In this appendix we want to define the quasimomentum
$\pu(\uu)$ in terms of a suitable, elementary elliptic function
\begin{equation}
r(\uu):=\frac{\theta_1\left(\uu+\frac14\right)}{\theta_1\left(\uu-\frac14\right)}\equiv
  \re^{\i \pu(\uu)},\label{rmomentum}
\end{equation}
and to express all other
elliptic functions, especially the ratio of $\theta_3$ functions in (\ref{BetheState}) 
and the energy (\ref{eq:XYenergy})
\begin{equation}
R(\uu):=\frac{\theta_3\left(\uu+\frac14\right)}{\theta_3\left(\uu-\frac14\right)},\qquad
{\cal E}(\uu):=2\frac{\theta_2(0)}{\theta'_1(0)} \left[\frac{\theta'_1\left(\uu-\frac14\right)}{\theta_1\left(\uu-\frac14\right)}
-  \frac{\theta'_1\left(\uu+\frac14\right)}{\theta_1\left(\uu+\frac14\right)}\right],\label{appenerg}
\end{equation}
in explicit terms of $\pu(\uu)$ thus avoiding the use of elliptic functions in
``practical calculations''.

We use several properties of the function $r(\uu)$
\begin{align}
  r\left(\uu+\tfrac12\right)=-\frac1{r(\uu)},\qquad  r\left(\uu+\tau\right)=-{r(\uu)},\qquad
  r\left(-\uu\right)=\frac1{r(\uu)},\qquad
r\left(\uu+\tfrac12+\tau\right)=\frac1{r(\uu)},\label{rident}
\end{align}
which follow more or less directly from the definition and elementary
properties of the $\theta$-functions. We have several special values of $r(\uu)$
\begin{align}
&\qquad r(0)=-1,\quad r\left(\pm\tfrac\tau2\right)=\mp\i,\quad r(\tau)=1,\qquad\qquad\label{unitcirc}\\
&\pm r\left(\tfrac14\pm\tfrac\tau2\right)=-\frac\i{J_y},\quad
\pm r\left(-\tfrac14\pm\tfrac\tau2\right)=-\i{J_y},\quad r'\left(\pm\tfrac14\pm\tfrac\tau2\right)=0,
\end{align}
where the identities in the second line follow from the symmetry
$r\left(\uu+\tfrac12+\tau\right)={r(-\uu)}$(see (\ref{rident})) and
\begin{equation}
J_y=\frac1{J_x}=\frac{\theta_4(0)}{\theta_3(0)}=-\i\frac{\theta_1\left(\frac\tau2\right)}{\theta_1\left(\frac12+\frac\tau2\right)}.
\end{equation}

\noindent
{\bf Expression of $R(\uu)$ in terms of $r(\uu)$}\\
We have the identity
\begin{equation}
R(\uu)^2=\frac{r(\uu)^2J_y^2+1}{r(\uu)^2+J_y^2},
\end{equation}
which is proven by use of Liouville's theorem. We note that both sides
are meromorphic functions and doubly periodic with periods $1, \tau$.  The two
sides have same poles and zeros and evaluate to 1 for argument $\uu=0$. In detail:
the poles of $r(\uu)$ on the RHS cancel in the ratio. The only poles of the
RHS are the zeros of the denominator. The only pole of the LHS in the period
rectangle (width 1, height $\tau$) is $-\tfrac14-\tfrac\tau2$ and of degree
2. This point is a zero of the denominator of the RHS, and it is a zero of 2nd
order due to the vanishing of the derivative of $r(\uu)$ at this point (the
second derivative can not be zero). In a similar way we see that
$\tfrac14-\tfrac\tau2$ is a zero of second order of LHS
and RHS. Special value of LHS and RHS at $u=0$: use $R(0)=+1$, $r(0)=-1$.\\

Now we see that $R(\uu)$ can be calculated from $r(\uu)=\re^{\i \pu}$
\begin{equation}
  R(\uu)=\mp\sqrt{\frac{J_y^2+1/r(\uu)^2}{J_y^2+r(\uu)^2}}\,r(\uu)\quad\hbox{for}\ J_y^2>1\quad\hbox{and}\quad
  =\mp\sqrt{\frac{1+J_y^2r(\uu)^2}{1+J_y^2/r(\uu)^2}}\,\frac1{r(\uu)}\quad\hbox{for}\ J_y^2<1,
\label{eqRfromr}
\end{equation}
where the sign $-$ ($+$) holds for $\uu\in\i {\mathbb R}$
($\uu\in\i {\mathbb R}+1/2$) and the square root is the standard one with
branch cut along $(-\infty,0]$. (In the main text, e.g.\ Fig.~\ref{Fig:disprels}
the second case in (\ref{eqRfromr}) applies.) Note that the argument of the
square root lies on the unit circle and never hits $-1$. The signs are then
fixed by considering the
points $\uu=0$ and $\uu=1/2$ yielding $r(0)=-1$, $R(0)=+1$ and $r(1/2)=R(1/2)=+1$.\\

\noindent
{\bf The dispersion relation: $\E(\uu)$ in terms of $r(\uu)$}\\
We have the identity
\begin{equation}
  -\left[\frac{\theta'_1\left(0\right)}{\theta_2\left(0\right)}
    \frac{\theta_1\left(\frac\tau2\right)}{\theta_1\left(\frac12+\frac\tau2\right)}\right]^2\, \left[r(\uu)^2+\frac1{J_y^2}\right]
  \left[\frac1{r(\uu)^2}+\frac1{J_y^2}\right]
  =\left[\frac{\theta'_1\left(\uu-\frac14\right)}{\theta_1\left(\uu-\frac14\right)}
-  \frac{\theta'_1\left(\uu+\frac14\right)}{\theta_1\left(\uu+\frac14\right)}\right]^2,\label{mainident}
\end{equation}
which is proven by using Liouville's theorem. We
note that both sides are meromorphic functions and doubly periodic with
periods $1, \tau$.  For $r(\uu)$ the shift of $u$ by $\tau$ leads to a minus
sign, but that drops out on the LHS. The shift by $\tau$ in the argument
leaves the RHS invariant as the
logarithmic derivatives of $\theta_1$ yield the same cancelling additive
constant
\begin{equation}
  \frac{\theta'_1\left(\uu+\tau\right)}{\theta_1\left(\uu+\tau\right)}
  =-2\pi \i+\frac{\theta'_1\left(\uu\right)}{\theta_1\left(\uu\right)}.\label{logdertau}
\end{equation}
Obviously, both sides have the same two poles per period
rectangle, e.g. at $u=\pm 1/4$, each of order 2. Both sides have the same
zeros of order 2 at $u=\pm 1/4+\tau/2$. The zeros deserve more explanations:
Inserting in (\ref{logdertau}) $u=-\tau/2$ and using that $\theta_1$ is an odd
function we get
\begin{equation}
  \frac{\theta'_1\left(\frac{\tau}{2}\right)}{\theta_1\left(\frac{\tau}{2}\right)}
  =-2\pi\i+\frac{\theta'_1\left(-\frac{\tau}{2}\right)}{\theta_1\left(-\frac{\tau}{2}\right)}
  =-2\pi\i-\frac{\theta'_1\left(\frac{\tau}{2}\right)}{\theta_1\left(\frac{\tau}{2}\right)}.
  \label{logdertau1}
\end{equation}
Inserting in (\ref{logdertau}) $u=1/2-\tau/2$ and using that $\theta_1$ is an
odd function with period 1 we get
\begin{equation}
  \frac{\theta'_1\left(\frac{1+\tau}{2}\right)}{\theta_1\left(\frac{1+\tau}{2}\right)}
  =-2\pi\i+\frac{\theta'_1\left(\frac{1-\tau}{2}\right)}{\theta_1\left(\frac{1-\tau}{2}\right)}
  =-2\pi\i-\frac{\theta'_1\left(\frac{1+\tau}{2}\right)}{\theta_1\left(\frac{1+\tau}{2}\right)}.
  \label{logdertau2}
\end{equation}
So we get
\begin{equation}
  \frac{\theta'_1\left(\frac{\tau}{2}\right)}{\theta_1\left(\frac{\tau}{2}\right)}
  =  \frac{\theta'_1\left(\frac{1+\tau}{2}\right)}{\theta_1\left(\frac{1+\tau}{2}\right)}
  =-\pi\i.
  \label{logdertau3}
\end{equation}
Therefore $u=1/4+\tau/2$ is a zero of the RHS of (\ref{mainident}) and clearly
of order 2. And the same holds for $u=-1/4+\tau/2$.

The two points $u=\pm1/4+\tau/2$ are clearly also zeros of the LHS and of
order 2. This is because the derivative of $r(\uu)$ at these points is zero
(and the second derivative has to be non-zero).

Now we know that the ratio of LHS and RHS has to be constant,
which can be obtained from evaluating the leading behaviour at the pole at $u=1/4$.

Now we obtain the expression for the energy ${\cal E}(\uu)$ by multiplying
(\ref{mainident}) with the square of $2{\theta_2(0)}/{\theta'_1(0)}$
\begin{align}
\frac{{\cal  E}^2(\uu)}{4}&=J_y^2\left[r(\uu)^2+\frac1{J_y^2}\right]\left[\frac1{r(\uu)^2}+\frac1{J_y^2}\right]
=r(\uu)^2+\frac1{r(\uu)^2}+{J_y^2}+\frac1{J_y^2}\no\\
&=\left[r(\uu)+\frac1{r(\uu)}\right]^2+\left[J_y-\frac1J_y\right]^2=4\cos^2\pu(\uu)+(J_y-J_x)^2,
\end{align}
from which follows
\begin{equation}
{\cal  E}(\uu)=\mp2\sqrt{4\cos^2\pu(\uu)+(J_y-J_x)^2},\label{appenerg2}
\end{equation}
where the sign $-$ ($+$) holds for $\uu\in\i {\mathbb R}$ ($\uu\in\i {\mathbb
  R}+1/2$). The sign is fixed by realizing ${\cal  E}(0)>0$ and ${\cal
  E}(\uu+1/2)=-{\cal  E}(\uu)$. Alternatively, with the definition of 
the invertible momentum function $\pu(\uu)$ (see below) we find $-$ holds for
$\pu\in(\pi/2,3\pi/2]$ and $+$ for $\pu\in(-\pi/2,\pi/2]$.\\

Finally we arrive at the following identity
\begin{equation}
\i\left[\frac{1}{r(\uu)R^{\pm 1}(\uu)}-{r(\uu)}{R^{\pm 1}(\uu)}\right]
=4 J_{y}^{\pm1}\,\frac{\sin [2p(\uu)]}{{\cal  E}(\uu)}
\quad =: \mathcal{X}(\uu,\pm1),\label{magfct}
\end{equation}
where we define the right hand side as function $\mathcal{X}$.\\

\noindent
{\bf Lines on which $r(\uu)$ takes unimodular values and $\pu(\uu)$ is real}\\
We see directly that $r(\uu)$ takes values of absolute value 1 on the imaginary
axis and on the imaginary axis shifted by $1/2$. From (\ref{unitcirc}) we see
that $r(\uu)$ runs over the unit circle for $\uu\in[-\tau,\tau]$ starting at $+1$
for $\uu=-\tau$, being $+\i$ for $\uu=-\tau/2$, $-1$ for $\uu=0$, $-\i$ for
$\uu=\tau/2$ and ending at $+1$ for $\uu=+\tau$. Similarly the unit
circle is surrounded in positive sense for $\uu\in[\tau,-\tau]+1/2$, however
beginning with $-1$ at $\uu=\tau+1/2$ and then decreasing the
imaginary part of $\uu$.

In this way all points of the unit circle are visited in the period rectangle
(width 1, height $2\tau$) at least two times. The precise number of such
visits has to be 2 as the number of poles is 2. Hence the above described
motion of $r(\uu)$ around the unit circle has to be monotonous. If not, then
certain points would be visited at least 3 times which would result in a
contradiction. Also we see that only on the imaginary axis and the imaginary
axis shifted by $1/2$ values on the
unit circle are taken.\\

\noindent
{\bf The quasimomentum $\pu(\uu)$ as invertible function}\\
Each value of the unit circle is taken by the function $r(\uu)$ at two points
in the period rectangle (width 1, height $2\tau$): this is for some imaginary
value $\uu$ and for the point $-\uu+1/2+\tau$. The inverse map from the unit
circle to the period rectangle is obviously not uniquely defined.

On the other hand, we have $r(\uu)=-r(\uu+\tau)$ and
$R(\uu)=-R(\uu+\tau)$. From this and (\ref{BetheState}) we obtain
\begin{equation}
\A_n(\vu)\,
R(\vu)^{1/2}\,r(\vu)=r(\vu)^{2m}\cdot
\begin{cases}\mp\sqrt{\frac{1+J_y^2r(\vu)^2}{1+J_y^2/r(\vu)^2}},&n=2m,\\
  1,&n=2m-1,
\end{cases}\label{amplstatewithr}
\end{equation}
where the sign $-$ ($+$) holds for $\uu\in\i {\mathbb R}$ ($\uu\in\i {\mathbb
  R}+1/2$). From (\ref{amplstatewithr}) we see that $\uu$ and
$\uu+\tau$ lead to the same state. This is ``consistent'' with the property
${\cal E}(\uu)={\cal E}(\uu+\tau)$. Therefore, we consider $\uu$ and
$\uu+\tau$ as equivalent. Now we consider solutions $u_j$ to the Bethe
equations on $\i {\mathbb R}$ reduced to the interval $(-\tau/2,+\tau/2]$, and
on $\i {\mathbb R}+1/2$ reduced to $(\tau/2,-\tau/2]+1/2$. The function $r(u)$
on the first interval parameterizes the semi-circle from $+\i$ over $-1$
to $-\i$ (first point excluded).  On the second interval it parameterizes the
semi-circle from $-\i$ over $+1$ to $+\i$ (first point excluded). Finally
we choose $\pu(\uu)$ to take values on the first interval out of
$(\pi/2,3\pi/2]$, and on the second interval out of $(-\pi/2,+\pi/2]$.\\

From (\ref{amplstatewithr}) we find
\begin{equation}
  \frac{\A_{2m}(\vu)}{\A_{2m-1}(\vu)}=
  \mp\sqrt{\frac{1+J_y^2r(\vu)^2}{1+J_y^2/r(\vu)^2}}=\mp\re^{\i
    [\pu(\uu)-2\varphi(p)]},\quad\hbox{where} \quad
  \tan(2\varphi)=\frac{J_x-J_y}{J_x+J_y}\tan p.\label{amplwithphi}
\end{equation}
\section{Magnetization of the XY model}\label{app;one-point}
For the one-point correlation $\langle \sigma_{n}^{x,y,z}\rangle$, we observe that 
\begin{align}
  &\bra{\Psi_M(\vvu)}\sigma_{n}^{x,y}\ket{\Psi_M(\vvu)}=0,\label{sigma;xy}\\
  &\bra{\Psi_M(\vvu)}\sigma_{n}^z\ket{\Psi_M(\vvu)}=\frac{1}{N}\sum_{\ik=1}^M
\mathcal{X}(\vu_\ik,(-1)^n),
    \label{sigma;z1}
\end{align}
where $\mathcal{X}(\vu,\pm1)$ was defined in (\ref{magfct}).

\begin{proof}
The state $\ket{\Psi_M(\vvu)}$ can be rewritten as
\begin{align}
|\Psi_M(\vvu)\rangle=O
\sum_{\mathbf{n}}\chi_{\mathbf{n}}(\vvu)\ket{1;\nn},
\end{align}
where 
\begin{align}
O=\mathbb{I}+(-1)^{\xi}\sigma_1^z\sigma_2^z\cdots\sigma_N^z,\quad\xi=\pm1.\label{O}
\end{align}
With the help of the operator $O$, we can easily prove that 
\begin{align}\label{xy;0}
\begin{aligned}
O^\dagger\,\sigma_n^x\,O&=
O^\dagger\,\sigma_n^y\,O&=0.
\end{aligned}
\end{align}
Therefore, one readily gets
\begin{align}
\bra{\Psi_M(\vvu)}\sigma_n^x\ket{\Psi_M(\vvu)}=\bra{\Psi_M(\vvu)}\sigma_n^y\ket{\Psi_M(\vvu)}=0.
\end{align} 
From Eq.~(\ref{xy;0}), one can also prove that
\begin{align}
\left\langle\sigma_{j_1}^{\alpha_1}\sigma_{j_2}^{\alpha_2}\cdots\sigma_{j_{2k+1}}^{\alpha_{2k+1}}\right\rangle=0,\quad \alpha_{j_1},\ldots,\alpha_{j_{2k+1}}=x,y.
\end{align}

Define an operator $\barT$, which shifts the lattice to the left by one site. 
One can prove 
\begin{align}
&\barT^2\ket{\kappa;n}=(-\ir)^2\ket{-\kappa;n-2},\qquad n\geq 3,\no\\
&\barT^2\ket{\kappa;n}=(-\ir)^{N-2}\ket{\kappa;n-2+N},\quad n=1,2,\\
&\barT^2\ket{\Psi_1(\vu)}=(-\ir)^2(-1)^\xi\left[\frac{\ell{1}(\vu+\frac{1}{4})}{\ell{1}(\vu-\frac{1}{4})}\right]^{2}\ket{\Psi_1(\vu)},\no\\
&\bra{\Psi_1(\vu)}(\barT^2)^\dagger=\ir^2(-1)^\xi\left[\frac{\ell{1}(\vu-\frac{1}{4})}{\ell{1}(\vu+\frac{1}{4})}\right]^{2}\bra{\Psi_1(\vu)}.
\end{align}
One can therefore conclude
\begin{align}
\bra{\Psi_1(\vu)}B_{n}\ket{\Psi_1(\vu)}=\bra{\Psi_1(\vu)}B_{n+2}\ket{\Psi_1(\vu)},
\end{align} 
where $B$ is an arbitrary operator. 
Analogously, one gets the following general identities 
\begin{align}
&\barT^2\ket{\Psi_M(\vvu)}=(-1)^{M}(-1)^\xi\,\eE^{2\ir\sum_{\ik=1}^M \vup_\ik}\ket{\Psi_M(\vvu)},\label{def;T}\\
&\bra{\Psi_M(\vvu)}B_{n}\ket{\Psi_M(\vvu)}=\bra{\Psi_M(\vvu)}B_{n+2}\ket{\Psi_M(\vvu)}.\label{Binary}
\end{align}
One can also check 
\begin{align}
\barT^N\ket{\Psi_M(\vvu)}&=(-1)^{\frac{MN}{2}}(-1)^{\frac{N\xi}{2}}\,\eE^{N\ir\sum_{\ik=1}^M \vup_\ik}\ket{\Psi_M(\vvu)}\no\\
&=(-1)^{\frac{M(N+2)}{2}}(-1)^{(\frac{N}{2}+M)\xi}\ket{\Psi_M(\vvu)}\no\\
&=\ket{\Psi_M(\vvu)}.
\end{align}

Due to the binary property of $\langle \sigma_{n}^{z}\rangle$, we only need to calculate the quantity $\langle \sigma_{N}^{z}\rangle$ and $\langle \sigma_{N-1}^{z}\rangle$.  
When $M=1$, we have 
\begin{align}
&\sigma_N^z\ket{\kappa;n}=\ir^{N-1}\ket{\kappa;n,N-1},\quad n\leq N-2,\no\\
&\sigma_N^z\ket{\kappa;N-1}=\ir\ket{\kappa;N},\no\\
&\sigma_N^z\ket{\kappa;N}=-\ir\ket{\kappa;N-1}.
\end{align}
Therefore, we get 
\begin{align}
\bra{\Psi_1(\vu)}\sigma_N^z\ket{\Psi_1(\vu)}&=\frac{\ir}{2N}\sum_{\kappa=\pm1}\left[\A_{N}(\vu)^*\A_{N-1}(\vu)\braket{\kappa;N}{\kappa;N}-\A_{N-1}(\vu)^*\A_{N}(\vu)\braket{\kappa;N-1}{\kappa;N-1}\right]\no\\
&=\frac{\ir}{N}\left[\frac{\ell{1}(\vu-\frac{1}{4})}{\ell{1}(\vu+\frac{1}{4})}\frac{\ell{3}(\vu-\frac{1}{4})}{\ell{3}(\vu+\frac{1}{4})}-\frac{\ell{1}(\vu+\frac{1}{4})}{\ell{1}(\vu-\frac{1}{4})}\frac{\ell{3}(\vu+\frac{1}{4})}{\ell{3}(\vu-\frac{1}{4})}\right]=\frac{1}{N}\,\mathcal{X}(\vu,+1).
\end{align}
Using the same method, we have
\begin{align}
\begin{aligned}
&\sigma_{N-1}^z\ket{\kappa;n}=\ir^{N-1}\ket{\kappa;n,N-1},\quad n\leq N-3,\\
&\sigma_{N-1}^z\ket{\kappa;N-2}=\ir\ket{\kappa;N-1},\\
&\sigma_{N-1}^z\ket{\kappa;N-1}=-\ir\ket{\kappa;N-2},\\
&\sigma_{N-1}^z\ket{\kappa;N}=\ir^{N-3}\ket{\kappa;N-2;N-1},
\end{aligned}
\end{align}
and 
\begin{align}
  \bra{\Psi_1(\vu)}\sigma_{N-1}^z\ket{\Psi_1(\vu)}=\frac{\ir}{N}\left[\frac{\ell{1}(\vu-\frac{1}{4})}{\ell{1}(\vu+\frac{1}{4})}\frac{\ell{3}(\vu+\frac{1}{4})}{\ell{3}(\vu-\frac{1}{4})}-\frac{\ell{1}(\vu+\frac{1}{4})}{\ell{1}(\vu-\frac{1}{4})}\frac{\ell{3}(\vu-\frac{1}{4})}{\ell{3}(\vu+\frac{1}{4})}\right]
=\frac{1}{N}\,\mathcal{X}(\vu,-1).
\end{align}
For $M=2$ case, the situation is slightly different. Let's first consider $\bra{\Psi_2(\vvu)}\sigma_N^z\ket{\Psi_2(\vvu)}$ as an example. We know 
\begin{align}
\begin{aligned}
&\sigma_N^z\ket{\kappa;n_1,n_2}=\ir^{N-1}\ket{\kappa;n_1,n_2,N-1},\quad n_1<n_2\leq N-2,\\
&\sigma_N^z\ket{\kappa;n_1,N-1}=\ir\ket{\kappa;n_1,N},\quad n_1\leq N-2,\\
&\sigma_N^z\ket{\kappa;n_1,N}=-\ir\ket{\kappa;n_1,N-1},\quad n_1\leq N-2,\\
&\sigma_N^z\ket{\kappa;N-1,N}=(-\ir)^{N-1}\ket{\kappa;N}.
\end{aligned}
\end{align}
Then, one can prove 
\begin{align}
\bra{\Psi_2(\vvu)}\sigma_N^z\ket{\Psi_2(\vvu)}&=2\ir\sum_{n=1}^{N-2}\left[\chi_{n,N}(\vvu)^*\chi_{n,N-1}(\vvu)-\chi_{n,N-1}(\vvu)^*\chi_{n,N}(\vvu)\right]\no\\
&=2\ir\sum_{n=1}^{N}\left[\chi_{n,N}(\vvu)^*\chi_{n,N-1}(\vvu)-\chi_{n,N-1}(\vvu)^*\chi_{n,N}(\vvu)\right]\no\\
&\overset{(\ref{xsqone})}{=}\frac{\ir}{N^2}\sum_{n=1}^N\sum_{j\neq \ik}\left[\A_n(\vu_j)^*\A_N(\vu_\ik)^*\A_n(\vu_j)\A_{N-1}(\vu_\ik)-\A_n(\vu_j)^*\A_{N-1}(\vu_\ik)^*\A_n(\vu_j)\A_{N}(\vu_\ik)\right]\no\\
&=\frac{1}{N}\sum_{\ik=1}^2\mathcal{X}(\vu_\ik,+1).\label{Sigmaz;M2}
\end{align}
Using the same technique, we can prove Eq.~(\ref{sigma;z1}).
\end{proof}

\section{Two-point correlations}\label{app;two-points}

Define the following two-point operator 
\begin{align}
X_n=\sigma_n^x\sigma_{n+1}^y-\sigma_n^y\sigma_{n+1}^x.\label{CurrentOperator}
\end{align}
One can prove
\begin{align}
&\bra{\Psi_M(\vvu)}\barT^\dagger\,X_n\,\barT\ket{\Psi_M(\vvu)}\no\\
&=\bra{\Psi_M(\vvu)}Y^\dagger X_nY\ket{\Psi_M(\vvu)}\no\\
&=\bra{\Psi_M(\vvu)} X_n\ket{\Psi_M(\vvu)},\label{Shift}
\end{align}
where \begin{align}
&Y=\frac{1}{\sqrt{2^N}}\prod_{\ik=1}^{N/2}(\sigma_{2\ik-1}^x+\sigma_{2\ik-1}^y)(\sigma_{2\ik}^x-\sigma_{2\ik}^y),\qquad Y^\dagger X_nY=X_n.\label{def;Y}
\end{align}
From Eqs.~(\ref{Binary}) and (\ref{Shift}), we thus conclude 
\begin{align}
\bra{\Psi_M(\vvu)}X_n\ket{\Psi_M(\vvu)}=\bra{\Psi_M(\vvu)}X_m\ket{\Psi_M(\vvu)}.
\end{align}
Without losing generality, we consider $\bra{\Psi_M(\vvu)}X_1\ket{\Psi_M(\vvu)}$. When $M=1$, we have 
\begin{align}
\begin{aligned}
&\sigma_1^x\sigma_{2}^y\ket{\kappa;n}=\ket{\kappa;n},\quad n\geq 2,\\
&\sigma_1^x\sigma_{2}^y\ket{\kappa;1}=-\ket{\kappa;1},\\
&\sigma_1^y\sigma_{2}^x\ket{\kappa;n}=\ir^2\ket{-\kappa;2,n},\quad 2<n<N,\\
&\sigma_1^y\sigma_{2}^x\ket{\kappa;N}=(-\ir)^{N-2}\ket{-\kappa;2},\\
&\sigma_1^y\sigma_{2}^x\ket{\kappa;2}=\ir^{N-2}\ket{-\kappa;N},\\
&\sigma_1^y\sigma_{2}^x\ket{\kappa;1}=-\ir^2\ket{-\kappa;1,2}.
\end{aligned}
\end{align} 
Then, we can get 
\begin{align}
&\bra{\Psi_1(\vu)}\sigma_1^x\sigma_{2}^y\ket{\Psi_1(\vu)}=1-\frac{2\A_1(\vu)^*\A_1(\vu)}{N}=1-\frac{2}{N},\\
&\bra{\Psi_1(\vu)}\sigma_1^y\sigma_{2}^x\ket{\Psi_1(\vu)}=\frac{(-1)^{\xi}}{N}[\A_2(\vu)^*\A_N(\vu)+\A_N(\vu)^*\A_2(\vu)]=-\frac{\mathcal{Y}(\vu)}{N}-\frac{2}{N},\\
&\bra{\Psi_1(\vu)} X_1\ket{\Psi_1(\vu)}=1+\frac{\mathcal{Y}(\vu)}{N},
\end{align} 
where 
\begin{align}
\mathcal{Y}(\vu)&=\left[\frac{\ell{1}(\vu-\frac{1}{4})}{\ell{1}(\vu+\frac{1}{4})}\right]^{2}+\left[\frac{\ell{1}(\vu+\frac{1}{4})}{\ell{1}(\vu-\frac{1}{4})}\right]^{2}-2.
\label{def:Y3}
\end{align}
For $M\geq 2$ cases, we can use Eq. (\ref{xsqone}) to prove
\begin{align}
&\bra{\Psi_M(\vvu)} \sigma_1^x\sigma_{2}^y\ket{\Psi_M(\vvu)}=1-\frac{2M}{N},\\
&\bra{\Psi_M(\vvu)}X_n\ket{\Psi_M(\vvu)}=1+\frac1N\sum_{\ik=1}^M\mathcal{Y}(\vu_\ik)=1-\frac{4}{N}\sum_{\ik=1}^M\sin^2\vup_j.
\end{align} 
We see that the two-point correlation $\bra{\Psi_M(\vvu)}X_n\ket{\Psi_M(\vvu)}$ in the XY model is the same as that in the XX model.

\section{The spin-helix eigenstate}
For an XY model with $N=4m,\,m\in\mathbb{N^+}$, we can construct the following states 
\begin{align}
\begin{aligned}
\ket{\Omega_1}&=\frac{1}{\sqrt{2^N}}\bigotimes_{n=1}^N\binom{1}{\ir^{n-1}},\\
\ket{\Omega_2}&=\frac{1}{\sqrt{2^N}}\bigotimes_{n=1}^N\binom{1}{\ir^{n+1}},\\
\ket{\Omega_3}&=\frac{1}{\sqrt{2}}(\ket{\Omega_1}+\ket{\Omega_2}),\\
\ket{\Omega_4}&=\frac{1}{\sqrt{2}}(\ket{\Omega_1}-\ket{\Omega_2}).
\end{aligned}
\end{align}
It can be proved that $\ket{\Omega_{1,2,3,4}}$ are all eigenstates of the Hamiltonian. The magnetization profiles in the aforementioned states are 
\begin{align}
\begin{aligned}
&\langle\sigma_{2\ik}^x\rangle_{1,2}=0,\quad \langle\sigma_{2\ik+1}^x\rangle_{1}=(-1)^\ik,\quad \langle\sigma_{2\ik+1}^x\rangle_{2}=(-1)^{\ik+1},\\
& \langle\sigma_{2\ik+1}^y\rangle_{1,2}=0,\quad\langle\sigma_{2\ik}^y\rangle_{1}=(-1)^{\ik+1},\quad \langle\sigma_{2\ik}^y\rangle_{2}=(-1)^{\ik},\\
&\langle\sigma_{\ik}^z\rangle_{1,2,3,4}=\langle\sigma_{\ik}^x\rangle_{3,4}=\langle\sigma_{\ik}^y\rangle_{3,4}=0.
\end{aligned}
\end{align}
One can also get the two-point correlation under the states $\ket{\Omega_{1,2,3,4}}$ 
\begin{align}
\langle X_{\ik}\rangle_{1,2,3,4}=1.
\end{align} 

\section{XX limit}\label{app;XX-limit}
Since the chiral basis vector $\ket{\pm1;\nn}$ is the same for both the XX model and the XY model, all the calculation techniques domesticated in the appendices also work for the XX model. The only difference is that the binary factors $\left[\frac{\ell{3}(\vu_\ik+\frac{1}{4})} {\ell{3}(\vu_\ik-\frac{1}{4})}\right]^{\pm\frac12}$ exist in the XY chain, while they will disappear in the XX case.

It should be noted that the direct reduction of the XY model to the XX model may yield results different from those presented in Table \ref{Table2}.

When $\tau\to+\ir\infty$, the XY model degenerates into XX model with $J_{x,y}\to 1$. For a large $\tau$, most of the Bethe roots are finite. In contrast, two Bethe roots exhibit significantly large values 
\begin{align}
\vu_\ik=\frac{\tau}{2},\,\frac{1-\tau}{2}.
\end{align}
In the XX limit, for the finite Bethe roots, we see that 
\begin{align}
\ell{3}(\vu_\ik\pm\tfrac14)=1,\quad \A_{n}(\vu_\ik)
&=\eE^{\ir n\vup_\ik},
\end{align}
which is consistent with Theorem \ref{thm2}. The quantities  $\mathcal{X}(\vu_\ik, \pm 1)$ for finite $\vu_\ik$ read 
\begin{align}
\mathcal{X}(\vu_\ik,+1)=\mathcal{X}(\vu_\ik,-1)=\ir(\eE^{-\ir\vup_\ik}-\eE^{\ir\vup_\ik})=2\sin\vup_\ik.
\end{align} 
For the remaining two infinite roots: $\vu_\ik=\frac{\tau}{2},\,\frac{1-\tau}{2}$, the function $\A_{n}(\vu_\ik)$ has the following form
\begin{align}
\A_{n}(\tfrac{\tau}{2})=\frac{(-\ir)^{n}- \ir^{n+1}}{\sqrt{2}},\quad 
\A_{n}(\tfrac{1-\tau}{2})=\frac{\ir^{n}+ (-\ir)^{n+1}}{\sqrt{2}}.
\end{align}  
In the limit $\tau\to+\ir\infty$, the direct reduction of the chiral Bethe state in Theorem \ref{thm3} will yield recombined eigenstates instead of the chiral Bethe states in Theorem \ref{thm2} when  $\frac{\tau}{2}$ or $\frac{1-\tau}{2}$ is included in the Bethe root set $\{\vu_1,\ldots,\vu_M\}$.
One can also prove that 
\begin{align}
&\mathcal{X}(\tfrac{\tau}{2},\pm1)=\mathcal{X}(\tfrac{1-\tau}{2},\pm1)=0,\quad 
\mathcal{Y}(\tfrac{\tau}{2})=\mathcal{Y}(\tfrac{1-\tau}{2})=-4.
\end{align}

\section{Proof of (\ref{eq:JvarianceExact})}
\label{App;Current}
We introduce the fermionic partition sum 
\begin{align}
&Z= \prod_p \left(  1+ \eE^{h j_p + \mu} \right),
\end{align}
where $\mu$ and $h$ are chemical potentials for the ``particle number" $M$ and the current $J$ respectively. 
Denoting $\Phi=\ln Z$ and differentiating w.r.t.~$\mu$ we get a relation between $M$ and $\mu$, namely,
\begin{align}
&M =\frac{\partial \Phi}{\partial \mu}\Big|_{h=0}=  N \partial_\mu \ln(  1+ \eE^{\mu}),
\end{align}
leading to 
\begin{align}
&\frac{M}{N} = \frac{1} {  1+ \eE^{- \mu}}. \label{app:M(nu)}.
\end{align}
The average current $\bar\currJ$ is given analogously by differentiating $\Phi$ w.r.t.~$h$,  using 
$j_p = 2-8 \sin^2 p$ (see  last line of Table I) and substituting (\ref{app:M(nu)}), leading to 
\begin{align}
&\bar\currJ = 2N-4M.
\end{align}
Finally, to obtain the variance of the current for fixed $M$, we need 
$\partial^2_ h \Phi$ for fixed $M$, or $\partial_h J$ for fixed $M$.
This is obtained via use of standard thermodynamic relations 
\begin{align}
  &
\langle(J-\bar J)^2\rangle_M = \left( \frac{\partial J}
{\partial h}\right)_M =  \frac{\partial (J,M)}
{\partial (h,M)} =  
\frac{\partial (J,M)}
{\partial (h,\mu)}  \frac{\partial (h,\mu)}
{\partial (h,M)} \\
&=\left[  
\left( \frac{\partial J}{\partial h}\right)_\mu
 \left( \frac{\partial M}{\partial \mu}\right)_h -  \left( \frac{\partial J}{\partial \mu}\right)_h
 \left( \frac{\partial M}{\partial h}\right)_\mu
\right] \left( \frac{\partial \mu}{\partial M}\right)_h\\
&= \partial^2_h \Phi - \frac{(\partial_h  \partial_\mu \Phi)^2}{\partial^2_h \Phi}.
\end{align}
After some algebra,  we obtain the thermodynamic result
\begin{align}
  &
\langle(J-\bar J)^2\rangle_M =8 M \left( 1 - \frac{M}{N}\right). \label{app:resAndreas}
\end{align}
which is valid in the thermodynamic limit of large $N$ and fixed $M/N$.   However finite-size
corrections exist as shows a comparison of  (\ref{app:resAndreas}) with the exact
formula for $M=1$
\begin{equation}
\langle(J-\bar J)^2\rangle_1=\langle(-8\sin^2p_1+4)^2\rangle
=4^2\langle\cos^22p_1\rangle=8,\label{app:DispIID1}
\end{equation}
suggesting a slight modification of (\ref{app:resAndreas})
  \begin{align}
\langle(J-\bar J)^2\rangle_M=8 {M}\frac{ 1 - \frac{M}{N}}{ 1 - \frac{1}{N}},
\end{align}
i.e. (\ref{eq:JvarianceExact}).  Comparison of the above with exact numerics for
sizes $N\leq 24$ and arbitrary $M$ strongly suggests that
(\ref{eq:JvarianceExact}) provides the exact expression for the variance of
the current for any $M, N$.

For the $S^z$ eigenbasis, with $j_k = 8 \sin k$, the analogous investigation
yields the average current and variance for fixed magnetization $S^z=M-N/2$
\begin{equation}
\bar J=0,\qquad \langle(J-\bar J)^2\rangle_M=32 {M}\frac{ 1 - \frac{M}{N}}{ 1 - \frac{1}{N}}.
\label{eq:JvarianceExact2}
\end{equation}
\end{document}